\documentclass[preprint,12pt,nofootinbib]{revtex4}

\usepackage[]{graphicx}\graphicspath{{graph/}}
\usepackage{amsmath,amssymb,amsfonts}
\usepackage{epstopdf}
\usepackage{hyperref}
\hypersetup{
  colorlinks,
  citecolor=magenta,
  linkcolor=blue}
\usepackage{fancyhdr}
\usepackage{datetime}

\renewcommand{\Re}{{\rm Re\,}}

\newcommand{\p}{{\partial}}

\newcommand{\intinf}{\int_{-\infty}^\infty}
\makeatletter
\newcommand{\lambdabarb}{{\mathchoice
  {\smash@bar\textfont\displaystyle{0.25}{1.2}\lambda}
  {\smash@bar\textfont\textstyle{0.25}{1.2}\lambda}
  {\smash@bar\scriptfont\scriptstyle{0.25}{1.2}\lambda}
  {\smash@bar\scriptscriptfont\scriptscriptstyle{0.25}{1.2}\lambda}
}}
\newcommand{\smash@bar}[4]{%
  \smash{\rlap{\raisebox{-#3\fontdimen5#10}{$\m@th#2\mkern#4mu\mathchar'26$}}}%
}
\makeatother
\begin{document}


\title{Microbunched Electron Cooling with Amplification Cascades
 }

\author{G. Stupakov and P. Baxevanis}
\affiliation{SLAC National Accelerator Laboratory, Menlo Park, CA 94025}

\begin{center}
\end{center}

\begin{abstract}

The Microbunched Electron Cooling (MBEC) is a promising cooling technique that can find applications in future hadron and electron-ion colliders to counteract intrabeam scattering that limits the maximum achievable luminosity of the collider. To minimize the cooling time, one would use amplification cascades consisting of a drift section followed by a magnetic chicane. In this paper, we first derive and optimize the gain factor in an amplification section for a simplified one-dimensional model of the beam. We then deduce the cooling rate of a system with one and two amplification cascades. We also analyze the noise effects that counteract the cooling process through the energy diffusion in the hadron beam. Our analytical formulas are confirmed by numerical simulations for a set of model parameters.

\vfill
%
\end{abstract}

\maketitle

%
\section{Introduction}
%

\begin{figure}[htb]
\centering
\includegraphics[width=0.7\textwidth, trim=0mm 0mm 0mm 0mm, clip]{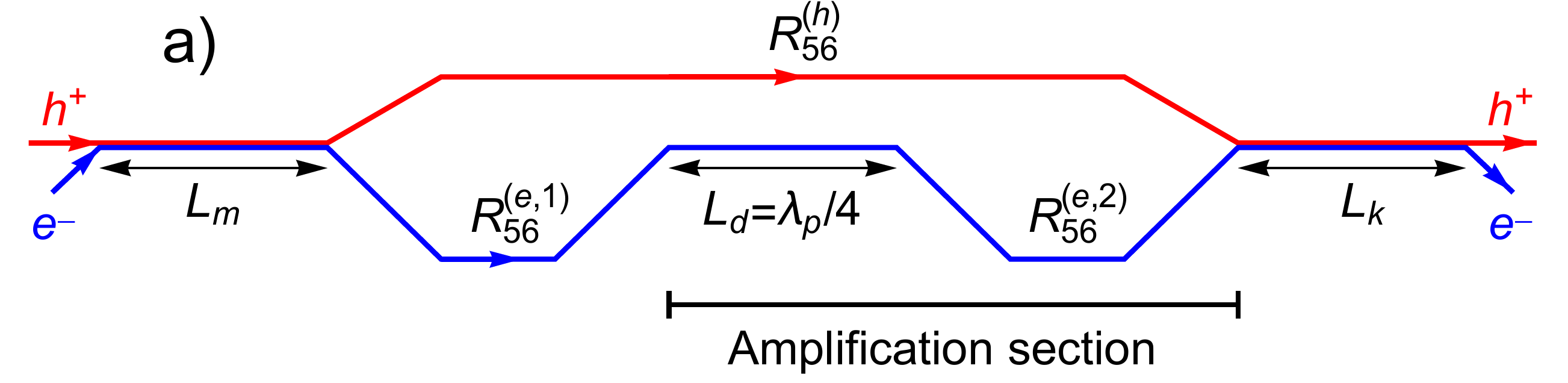}
\includegraphics[width=0.7\textwidth, trim=0mm 0mm 0mm 0mm, clip]{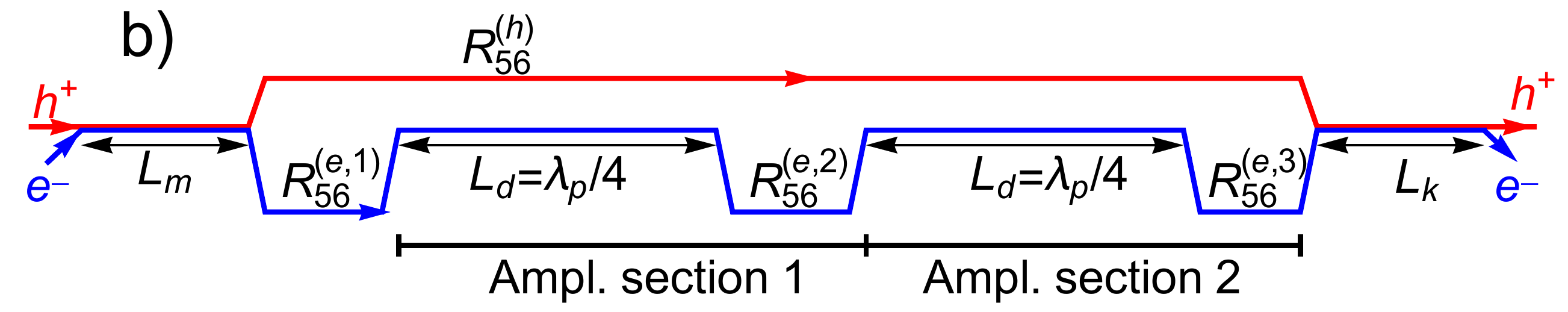}
\caption{Schematic of the microbunched electron cooling system: a) with one amplification section, b) with two amplification sections. Blue lines show the path of the electron beam, and the red lines indicate the trajectory of the hadron beam.}
\label{fig:1}
\end{figure}

Microbunched coherent electron cooling (MBEC) of relativistic hadron beams has been proposed by D. Ratner~\cite{Ratner2013} as a way to achieve cooling rates higher than those provided by the coherent cooling using a free electron laser~\cite{litvinenko09}. The mechanism of MBEC can be understood in a simple setup shown in Fig.~\ref{fig:1}. 
An electron beam with the same relativistic $\gamma$-factor as in the hadron beam, co-propagates with the hadrons in a section of length $L_m$ called the ``modulator''. In this section, the hadrons imprint microscopic energy perturbations onto the electrons via the Coulomb interaction. After the modulation, the electron beam passes through a dispersive chicane section, $R_{56}^{(e,1)}$, where the energy modulation of the electrons is transformed into a density fluctuation referred to as ``microbunching''. This chicane is followed by an amplification section consisting of a drift of length $L_d$ and another chicane $R_{56}^{(e,2)}$, as shown in Fig.~\ref{fig:1}a. If the length of the drift is equal to one-quarter of the plasma oscillation period in the electron beam, $\frac{1}{4}\lambda_p$, and the chicane strength is properly optimized, the density fluctuations in the electron beam generated by the chicane $R_{56}^{(e,1)}$ are increased in amplitude. This section can be repeated several times: Fig.~\ref{fig:1}b shows the setup with two amplification sections. Meanwhile, the hadron beam passes through its dispersive section, $R_{56}^{(h)}$, in which more energetic particles move in the forward direction with respect to their original positions in the beam, while the less energetic hadrons trail behind. When the beams are combined again in a section of length $L_k$, called the ``kicker'', the electric field of the induced density fluctuations in the electron beam acts back on the hadrons. With a proper choice of the chicane strengths, the energy change of the hadrons in the kicker leads, over many passages through the cooling section, to a gradual decrease of the energy spread of the hadron beam. The transverse cooling is achieved in the same scheme by introducing  dispersion in the kicker for the hadron beam.

Theoretical analysis of MBEC without amplification has been carried out in a recent study~\cite{Stupakov:2018anp}. In this paper, we extend the analysis of Ref.~\cite{Stupakov:2018anp} to include the amplification sections. Following the approach developed in Ref.~\cite{Stupakov:2018anp} we adopt a general framework in which we look at the dynamics of the fluctuations in both beams. We assume that before the beams start to interact, their density and energy fluctuations can be described as uncorrelated shot noise. In the process of interaction, the fluctuations in the electron and hadron beams establish correlations, and when the beams are recombined in the kicker the fluctuating electric field in the electron beam acts in a way that  decreases the energy spread in the hadron beam. As in Ref.~\cite{Stupakov:2018anp}, for the hadron-electron, as well as electron-electron interactions we adopt a model in which the particles are replaced by thin disks with a Gaussian transverse charge distribution.

The paper is organized as follows. In Sec.~\ref{sec:2} we summarize the Coulomb interaction between thin slices with transverse Gaussian distribution of charge. In Sec.~\ref{sec:3} we study plasma oscillations in a beam consisting of thin Gaussian slices. In Sec.~\ref{sec:4} we demonstrate that an initial sinusoidal modulation of small amplitude in a beam is amplified after the passage through a quarter of plasma wavelength drift and a subsequent chicane. The amplification factor derived in this section is then used, in Secs.~\ref{sec:5} and~\ref{sec:6}, for calculation of the cooling rate in an MBEC cooling system with one and two amplification sections, respectively. In Sec.~\ref{sec:6-1} we discuss the wake field associated with the amplified cooling which is related to the effective energy exchange of two hadrons located at a given distance $z$. In Sec.~\ref{sec:7} we present results of computer simulations of the cooling rates. In Sec.~\ref{sec:8} the noise and saturation effects in the cooling process are studied, and in Sec.~\ref{sec:9} numerical estimates of the hadron cooling are presented for the eRHIC electron-ion collider design. We concluded this paper with the summary in Sec.~\ref{sec:10}.

We use the Gaussian system of units throughout this paper.
%
\section{Interaction of charged Gaussian slices}\label{sec:2}
%

As was already mentioned in the Introduction, we treat the Coulomb interaction between  particles as if a hadron were a disk of charge $Ze$ with an axisymmetric Gaussian radial distribution with the rms transverse size equal to the rms transverse size of the beam. The electron is also modeled by a Gaussian disk of charge $-e$ with the same transverse profile. A similar Gaussian-to-Gaussian interaction model was used in 1D simulations of a longitudinal space charge amplifier in Ref.~\cite{dohlus2011sy}. 

In this model, a hadron of charge $Ze$ at the origin of the coordinate system exerts a force $f_z$  on an electron at coordinate $z$,
    \begin{align}\label{eq:1}
    f_z(z)
    =
    -
    \frac{Ze^2}{\Sigma^2}
    \Phi
    \left(\frac{z\gamma}{\Sigma}\right)
    ,
    \end{align}
where $\Sigma$ is the rms beam radius and the function $\Phi$ is defined by the following expression~\cite{geloni07ssy},
    \begin{align}\label{eq:2}
    \Phi(x)
    =
    \frac{1}{2}
    \left[
    \frac{x}{|x|}
    -
    \frac{x\sqrt{\pi}}{2}
    \exp
    \left(
    \frac{1}{4}x^2
    \right)
    \mathrm{erfc}
    \left(
    \frac{1}{2}|x|
    \right)
    \right]
    ,
    \end{align}
with erfc the complementary error function.  The function $\Phi$ is odd, $\Phi(-x) = -\Phi(x)$; its plot can be found in Ref.~\cite{Stupakov:2018anp}. Neglecting the relative longitudinal displacements of hadrons and electrons in the modulator, the force~\eqref{eq:1} causes the relative energy change $\Delta\eta$ of an electron located at coordinate $z$,
    \begin{align}\label{eq:3}
    \Delta\eta(z)
    =
    -
    \frac{Zr_eL_m}{\gamma \Sigma^2}
    \Phi
    \left(\frac{z\gamma}{\Sigma}\right)
    ,
    \end{align}
where $L_m$ is the length of the modulator and $r_e=e^2/m_ec^2$ is the classical electron radius, and we use the notation $\eta$ for the energy deviation $\Delta E$ of a particle normalized by the nominal beam energy $\gamma mc^2$, $\eta = \Delta E/\gamma mc^2$. Eq.~\eqref{eq:3} can also be considered as a Green function for the energy modulation of electrons induced by a delta-function density perturbation in the hadron beam.  

In our analysis we will assume that the beam radius in the amplification sections, $\Sigma_p$, may be different from the beam radius in the kicker and the modulator, $\Sigma$. For the electron-electron interaction in these sections we use Eq.~\eqref{eq:1} with $Z=-1$ and $\Sigma\to\Sigma_p$,
    \begin{align}\label{eq:4}
    F_z(z)
    =
    \frac{e^2}{\Sigma_p^2}
    \Phi
    \left(\frac{\gamma z}{\Sigma_p}\right)
    .
    \end{align}

\begin{figure}[htb]
\centering
\includegraphics[width=0.6\textwidth, trim=0mm 0mm 0mm 0mm, clip]{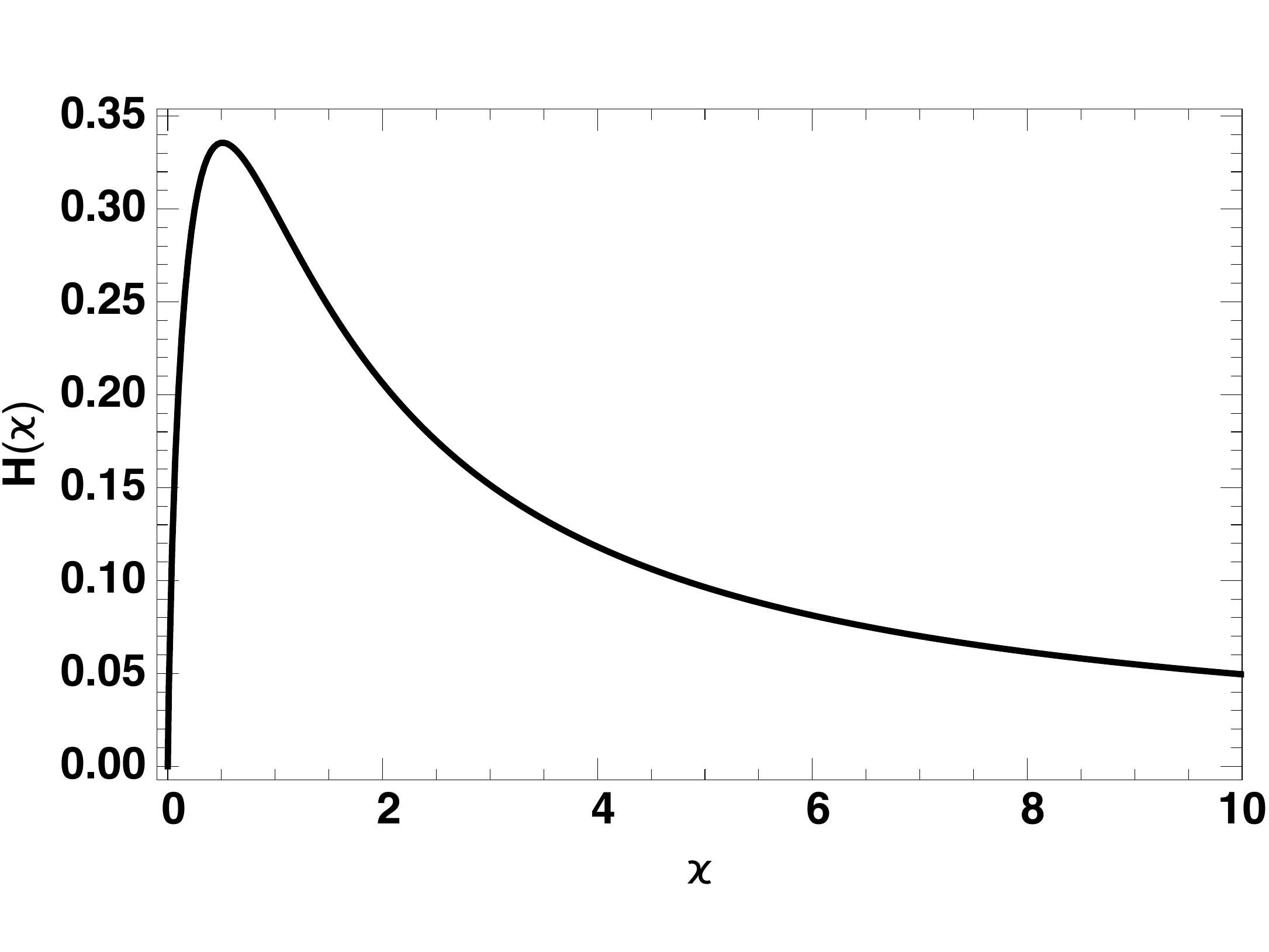}
\caption{Plot of function $H(\varkappa)$ for positive values of the argument.}
\label{fig:2}
\end{figure}

In what follows, we will also need the Fourier transform of the function $\Phi$. Because of the antisymmetry of the function $\Phi$ its Fourier transform is purely imaginary, so we define function $H$ as
    \begin{align}\label{eq:5}
    H(\varkappa)
    =
    \frac{i}{2}
    \intinf
    dx
    \Phi(x)
    e^{-i\varkappa x}
    =
    \int_0^\infty dx \Phi(x) \sin({\varkappa x})
    .
    \end{align}
The plot of function $H(\varkappa)$ is shown in Fig.~\ref{fig:2}.
For large values of the argument, $\varkappa\gg 1$, the function $H$ asymptotically approaches $1/2\varkappa$. One can also find an approximation for $H$ near the origin, $\varkappa\ll 1$, $H\approx \varkappa(2\ln\varkappa - \gamma_E)$, where $\gamma_E = 0.577$ is the Euler constant.

%
\section{Plasma oscillations in a Gaussian beam}\label{sec:3}
%

To calculate the increase in the amplitude of the fluctuations in the electron beam when it propagates through the amplification sections, we first need to analyze the beam plasma oscillations in the drift. A similar problem has been studied in Refs.~\cite{PhysRevSTAB.13.110703}, however, our treatment is simpler because we will assume a cold plasma and neglect the transverse degrees of freedom in the beam. In our analysis, we use the Vlasov equation for the distribution function in the longitudinal phase space, $f(z,\eta,t)$, normalized so that $\intinf f\,d\eta  = n$, with $n$ the number of particles in the beam per unit length. Here $z=s-vt$ is the longitudinal coordinate in the beam with $s$ the distance measured along the direction of the beam propagation in the lab frame, and $v$ the nominal beam velocity. We represent the distribution function as  $f= n_0 F_0(\eta)+\delta f(z,\eta,t)$, where $F_0(\eta)$ is the equilibrium beam distribution function, $n_0$ is the nominal linear particle density, and $\delta f$ describes small-amplitude time-dependent fluctuations in the beam, $|\delta f|\ll F_0$. We consider the fluctuations with the longitudinal scale much smaller than the bunch length and carry out our analysis in a small vicinity of a given location in the bunch where the variation of the distribution function $F_0$ with $z$ can be neglected; for this reason the coordinate $z$ is omitted from the arguments of the function $F_0$.

The linearized Vlasov equation for the perturbation of the distribution  function, $\delta f$, is
    \begin{align}\label{eq:6}
    \frac{\p \,\delta f}{\p t}
    +
    \frac{c\eta}{\gamma^2}
    \frac{\p \,\delta f}{\p z}
    +
    \dot \eta
    n_0F_0'(\eta)
    =
    0
    ,
    \end{align}
where $\dot \eta$ is the energy change per unit time. The rate of energy change is expressed through the longitudinal force in the electron beam,
    \begin{align}\label{eq:7}
    \dot\eta
    =
    \frac{1}{\gamma m_ec}
    \int_{-\infty}^\infty
    dz'
    \delta n(z',t)
    F_z(z-z')
    ,
    \end{align}
where 
    \begin{align}\label{eq:8}
    \delta n(z,t)
    =
    \intinf
    d\eta
    \,
    \delta f(z,\eta,t)
    ,
    \end{align}
and $F_z$ is given by Eq.~\eqref{eq:4}. Making the Fourier transform of Eqs.~\eqref{eq:6} and~\eqref{eq:7} and using the notations
    \begin{align}\label{eq:9}
    \delta \hat f_k(\eta,t)
    =
    \intinf
    dz\,
    e^{-ikz}
    \delta f(z,\eta,t)
    ,
    \qquad
    \delta \hat n_k(t)
    =
    \intinf
    dz\,
    e^{-ikz}
    \delta n(z,t)
    ,
    \end{align}
we obtain
    \begin{align}\label{eq:10}
    \frac{\p \,\delta \hat f_k}{\p t}
    +
    \frac{ikc\eta}{\gamma^2}
    \delta\hat f_k
    +
    \zeta(k)\,
    \delta \hat n_k
    n_0F_0'(\eta)
    =
    0
    ,
    \end{align}
with the effective impedance $\zeta(k)$ given by
    \begin{align}\label{eq:11}
    \zeta(k)
    =
    \frac{1}{\gamma mc}
    \int_{-\infty}^\infty
    d\xi
    e^{-ik\xi}
    F_z(\xi)
    =
    -
    \frac{2ie^2}{\Sigma_p\gamma^2 mc}
    H
    \left(\frac{k\Sigma_p}{\gamma}\right)
    ,
    \end{align}
where the function $H$ is defined by Eq.~\eqref{eq:5}. At large values of $k$, $k\gtrsim \gamma/\Sigma_p$, we have $H(\varkappa) \sim 1/\varkappa$, so the impedance in this region can be estimated as $\zeta\sim e^2/\Sigma_p^2\gamma m c k$.

\begin{figure}[htb]
\centering
\includegraphics[width=0.6\textwidth, trim=0mm 0mm 0mm 0mm, clip]{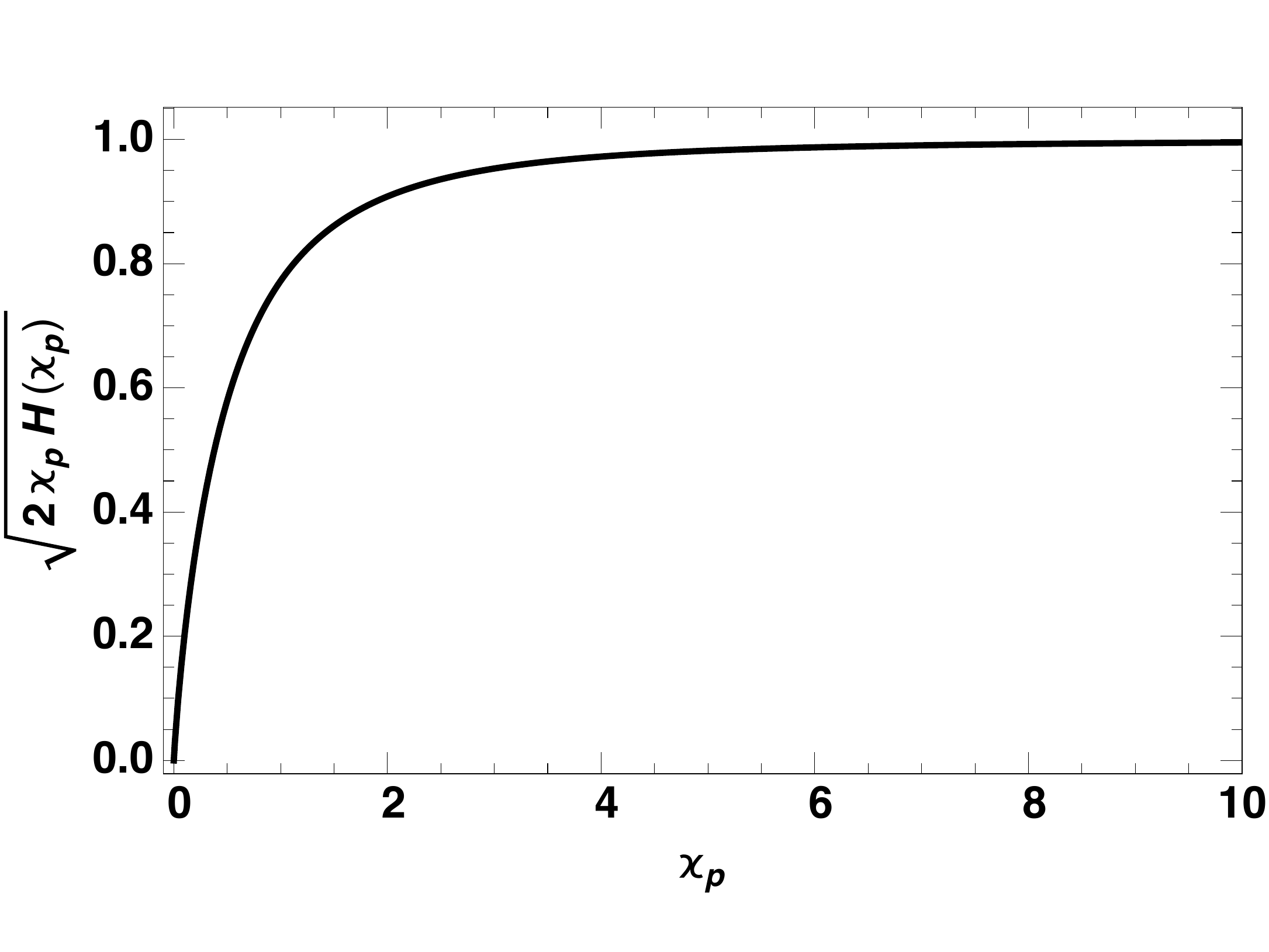}
\caption{Plot of the function $\sqrt{2\varkappa_p H(\varkappa_p)}$.}
\label{fig:3}
\end{figure}

In our analysis we will assume that the second term in Eq.~\eqref{eq:10} is much smaller than the third one. The conditions for such assumption are estimated at the end of this section. Neglecting the second term we can integrate the Vlasov equation over time,
    \begin{align}\label{eq:12}
    \delta\hat f_k(\eta,t)
    =
    \delta\hat f_k(\eta,0)
    -
    \zeta(k)\,
    n_0F_0'(\eta)
    \int_0^t
    dt'
    \delta \hat n_k(t')
    .
    \end{align}
To get an equation for $\delta \hat n_k(t)$ we integrate Eq.~\eqref{eq:10} over $\eta$,
    \begin{align}\label{eq:13}
    \frac{d \,\delta \hat n_k}{d t}
    +
    \frac{ikc}{\gamma^2}
    \delta \hat q_k
    =
    0
    ,
    \end{align}
where 
    \begin{align}\label{eq:14}
    \delta \hat q_k
    =
    \intinf
    d\eta
    \,\eta\,
    \delta \hat f_k
    \end{align}
is the averaged perturbation of the energy $\eta$ by $\delta \hat f_k$. 
Note that the large third term in Eq.~\eqref{eq:10} does not contribute to this equation, so we have to keep the contribution from the second term. We can also obtain an equation for $\delta \hat q_k$ by integrating Eq.~\eqref{eq:10} with weight $\eta$,
    \begin{align}\label{eq:15}
    \frac{d\,\delta \hat q_k}{d t}
    -
    \zeta(k)\,
    \delta \hat n_k
    n_0
    =
    0
    ,
    \end{align}
where we have neglected the contribution from the second term in Eq.~\eqref{eq:10}. Combining Eqs.~\eqref{eq:13} and~\eqref{eq:15} we find
    \begin{align}\label{eq:16}
    \frac{d^2 \,\delta \hat n_k}{d t^2}
    +
    \frac{ikc}{\gamma^2}
    \zeta(k)\,
    n_0
    \delta \hat n_k
    =
    0
    ,
    \end{align}
which is the equation for plasma oscillations in the beam. A somewhat different derivation of plasma oscillations in a relativistic beam is given in Ref.~\cite{PhysRevLett.102.154801}. Note that there is no Landau damping effects~\cite{PhysRevSTAB.13.110703} in this equation, which means that our assumption of the smallness of the second term in Eq.~\eqref{eq:10} is equivalent to the cold plasma approximation. It follows from Eq.~\eqref{eq:16} that the plasma frequency $\omega_p(k)$ is given by the following equation:
    \begin{align}\label{eq:17}
    \omega_p^2
    =
    \frac{ikcn_0}{\gamma^2}
    \zeta(k)
    =
    \frac{2kn_0e^2}{\Sigma_p\gamma^4 m}
    H
    \left(\frac{k\Sigma_p}{\gamma}\right)
    =
    2\,
    \Omega^2
    \varkappa_p
    H(\varkappa_p)
    ,
    \end{align}
with
    \begin{align}\label{eq:18}
    \Omega^2 = \frac{n_0e^2}{m\Sigma_p^2\gamma^3}
    \end{align}
and $\varkappa_p=k\Sigma_p/\gamma$. The plot of the function $\sqrt{2\varkappa_p H(\varkappa_p)}$ --- this function is equal to the normalized plasma frequency $\omega_p/\Omega$ ---  is shown in Fig.~\ref{fig:3}. 
One can see that the short-period plasma oscillations with $k\gg\gamma/\Sigma_p$,  the plasma frequency is approximately equal to $\Omega_p$, while for the long-wavelength oscillations $\omega_p$ decreases with the wavelength. 

We can now find the condition when the second  term in Eq.~\eqref{eq:10} is much smaller than the third one. In our estimates we will assume $k\sim \gamma/\Sigma_p$ which gives for the impedance $\zeta\sim e^2/\Sigma_p\gamma^2 m c$. Using the estimates $F_0'\sim 1/\sigma_e^2$, $\delta \hat f_k \sim \delta \hat n_k/\sigma_e$ and $\eta\sim\sigma_e$ we find that the ratio of the third term to the second one is approximately equal to
    \begin{align}\label{eq:19}
    \frac{e^2n_0}{\gamma m c^2\sigma_e^2}
    \sim
    \frac{1}{ \sigma_e^2}
    \frac{I_e}{\gamma I_A}
    ,
    \end{align}
which has to be much greater than one. Here $I_e = ecn_0$ is the peak beam current and $I_A=mc^3/e$ is the Alfv\'en current. In the next section, we will see that the same parameter~\eqref{eq:19} appears in the expression for the gain factor of the amplification section.

%
\section{Gain factor in an amplification cascade}\label{sec:4}
%

\begin{figure}[htb!]
\centering
\includegraphics[width=0.6\textwidth, trim=0mm 0mm 0mm 0mm, clip]{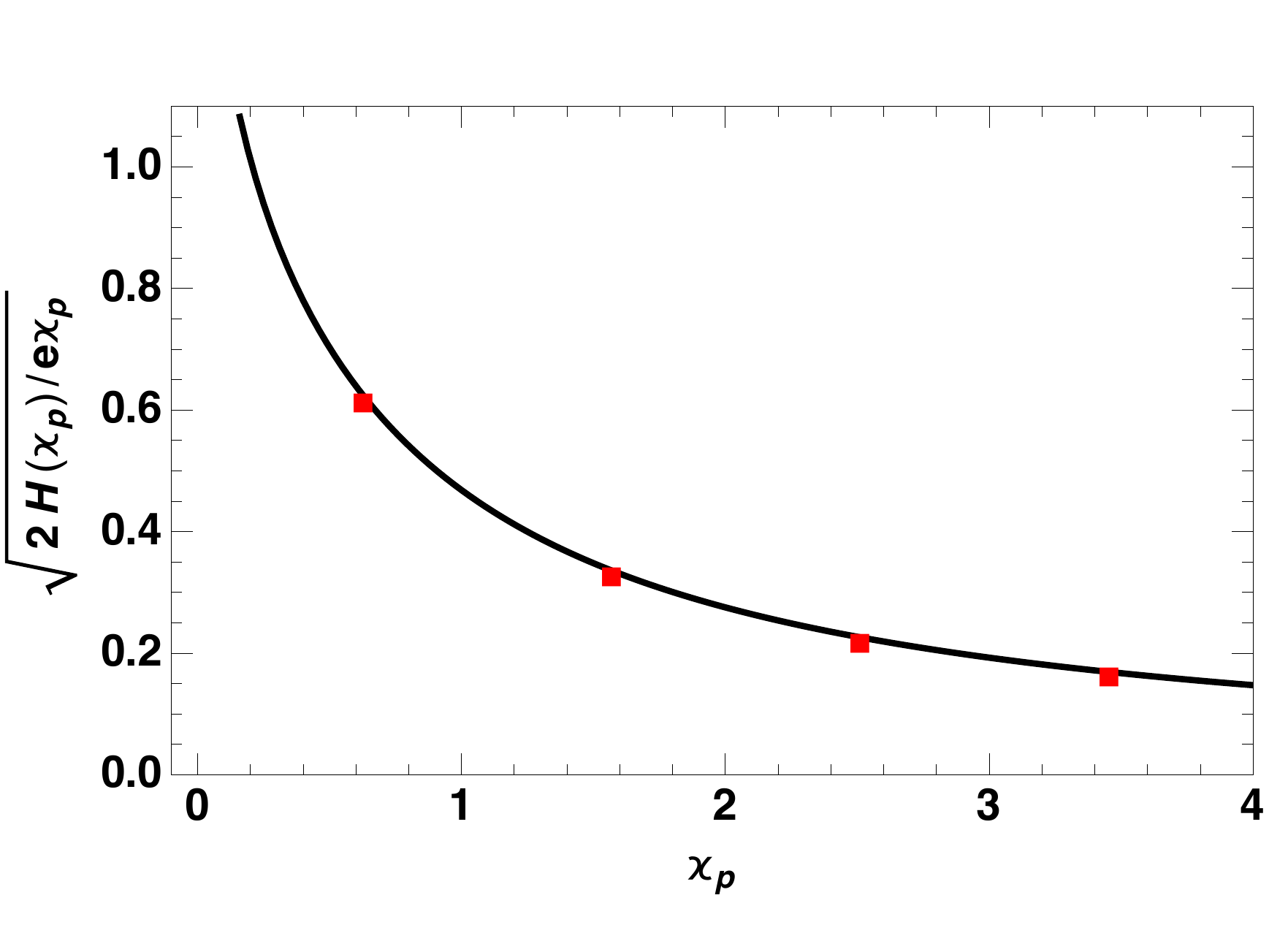}
\caption{Plot of the function $\sqrt{{2H(\varkappa_p)}/{\mathrm{e}\varkappa_p}}$ (solid curve). The dots show the result of the simulation of the amplification factor (see Section~\ref{sec:7}).}
\label{fig:4}
\end{figure}

The solution of Eqs.~\eqref{eq:13} and~\eqref{eq:15} is
    \begin{align}\label{eq:20}
    \delta \hat n_k
    =
    \delta \hat n_k(0)
    \cos(\omega_p t)
    -
    \frac{ikc}{\gamma^2\omega_p}
    \delta \hat q_k(0)
    \sin(\omega_p t)
    ,
    \end{align}
where $\delta \hat n_k(0)$ and $\delta \hat q_k(0)$ are the initial values of the density and energy perturbations in the beam. Let us compare the first and the second terms in this equation taking into account that plasma oscillations occur after the fist chicane, $R_{56}^{(e,1)}$, of the cooling section. In these estimates we again assume $k\sim \gamma/\Sigma_p$ and the optimal value for the chicane strength, $R_{56}^{(e,1)}\sim \Sigma_p/\gamma \sigma_e$ (see order of magnitude estimates in Ref.~\cite{Stupakov:2018HB}). The magnitude of $\delta \hat n_k(0)$ is estimated as $\delta \hat n_k(0)\sim kR_{56}^{(e,1)} \delta \hat q_k(0)\sim \delta \hat q_k(0)k\Sigma_p/\gamma \sigma_e $, so that the ratio of the second term to the first one in Eq.~\eqref{eq:20} is of the order of $c\sigma_e/\gamma\Omega\Sigma_p$. This combination of parameters turns out to be equal to the inverse of the square root of the parameter in Eq.~\eqref{eq:19}, 
    \begin{align}\label{eq:21}
    \frac{c\sigma_e}{\gamma\Omega\Sigma_p}
    \sim
    \sigma_e
    \sqrt{\frac{\gamma I_A}{I_e}}
    \ll
    1
    ,
    \end{align}
and, by assumptions, is much smaller than one. Hence we can neglect the second term in Eq.~\eqref{eq:20},
    \begin{align}\label{eq:22}
    \delta \hat n_k
    \approx
    \delta \hat n_k(0)
    \cos(\omega_p t)
    .
    \end{align}
Substituting this result in Eq.~\eqref{eq:12} we obtain
    \begin{align}\label{eq:23}
    \delta\hat f_k(\eta,t)
    =
    \delta\hat f_k(\eta,0)
    -
    \frac{1}{\omega_p}
    \zeta(k)\,
    n_0F_0'(\eta)
    \delta \hat n_k(0)
    \sin(\omega_p t)
    .
    \end{align}
Estimating the relative magnitude of the two terms on the right-hand side in this equation, as it was done in the derivation of Eq.~\eqref{eq:19}, we find that the ratio of the first term to the second one is given by the same parameter~\eqref{eq:21} and hence we can neglect the first term in Eq.~\eqref{eq:23}. Using this expression for $\delta\hat f_k$, we can find the linear density perturbation in the beam, $\delta \hat n_k^{(2)}$, after it passes through the second chicane  $R_{56}^{(e,2)}$ at the end of the drift (see Fig.~\ref{fig:1}a):
    \begin{align}\label{eq:24}
    \delta \hat n_k^{(2)}
    &=
    \intinf
    d\eta
    \delta\hat f_k
    e^{-ikR_{56}^{(e,2)}\eta}
    =
    -
    \frac{1}{\omega_p}
    \zeta(k)n_0
    \delta \hat n_k(0)
    g(k)
    \sin
    \left(
    \frac{\omega_p L_d}{c}
    \right)
    ,
    \end{align}
where we have replaced the time by the length of the drift divided by the speed of light, $t=L_d/c$, and
    \begin{align}\label{eq:25}
    g(k)
    =
    \intinf
    d\eta
    F_0'(\eta)
    e^{-ikR_{56}^{(e,2)}\eta}
    =
    ikR_{56}^{(e,2)}
    e^{-k^2(R_{56}^{(e,2)})^2\sigma_{e}^2/2}
    .
    \end{align}
The last expression in this formula is calculated for a Gaussian distribution function $F_0 = (2\pi)^{-1/2} \sigma_e^{-1} e^{-\eta^2/2\sigma_e^2}$ with $\sigma_e$ the rms relative energy spread in the electron beam. It makes sense to define the gain factor $G$ of the amplification section as $G=\delta \hat n_k^{(2)}/\delta \hat n_k(0)$. Using Eq.~\eqref{eq:11} for the impedance and Eq.~\eqref{eq:17} for the plasma frequency, after simple calculations, we find
    \begin{align}\label{eq:26}
    G
    =
    -
    \frac{1}{\sigma_e}
    \sqrt{\frac{2I_e}{\gamma I_A}}
    \sqrt{\varkappa_p    H\left(\varkappa_p\right)}\,
    q_p\,
    e^{-\varkappa_p^2q_p^2/2}
    \sin
    \left(
    \frac{\omega_p L_d}{c}
    \right)
    ,
    \end{align}
where $q_p = R_{56}^{(e,2)}\sigma_e\gamma/\Sigma_p$. The dependence of $G$ versus the transverse size  of the beam $\Sigma_p$ is mostly determined by the ratio $\sqrt{H({k\Sigma_p/\gamma})/\Sigma_p}$ and for a given value of $k$ this function, and hence the gain factor, increases when $\Sigma_p$ becomes smaller. As a function of the chicane strength, the gain factor reaches maximum at $q_p = 1/\varkappa_p$ with the maximum value of $G$ equal to
    \begin{align}\label{eq:27}
    G_\mathrm{max}
    &=
    -
    \frac{1}{\sigma_e}
    \sqrt{\frac{2I_e}{\gamma I_A}}
    \sqrt{\frac
    {2H(\varkappa_p)}
    {\mathrm{e}\varkappa_p}}
    \sin
    \left(
    \frac{\omega_p L_d}{c}
    \right)
    ,
    \end{align}
where `e'$\approx2.71$ is the base of the natural logarithm.  The plot of $\sqrt{{ 2H(\varkappa_p)}/    {\mathrm{e}\varkappa_p}}$ is shown in Fig.~\ref{fig:4}.

For a drift length equal to a quarter of the plasma wavelength, $L_d=\pi c/2\omega_p$, the sin function is equal to one. Note that in this case the gain factor~\eqref{eq:26} is negative (if $q_p>0$). It means that an amplification section also introduces a 180 degrees phase shift in harmonics of the plasma oscillations relative to their values at the beginning of the section.

%
\section{Cooling rate with one amplification section}\label{sec:5}
%

For the cooling time measured in the revolution periods in the ring, $N_\mathrm{c}$,  without the amplification sections (that is in the absence of the drift $L_d$ and the chicane $R_{56}^{(e,2)}$ in Fig.~\ref{fig:1}a), the following expression was derived in Ref.~\cite{Stupakov:2018anp}:
    \begin{align}\label{eq:28}
    N_\mathrm{c}^{-1}
    =
    \frac{2icr_hq_h}{\pi \Sigma \sigma_h}
    \Re
    \int_{0}^\infty
    {d\varkappa}
    {\cal Z}(\varkappa)
    \varkappa 
    e^{-\varkappa^2 q_h^2/2}
    ,
    \end{align}
where $\Sigma$ is the rms transverse beam size in the modulator and the kicker, $\sigma_h$ is the relative rms energy spread of the hadron beam, $r_h = (Ze)^2/m_hc^2$ with $m_h$ the hadron mass, $q_h$ is the normalized strength of the hadron chicane, $q_h = {R_{56}^{(h)}\sigma_{h}\gamma}/{\Sigma}$, and $\varkappa$ is the normalized wavenumber $k$, $\varkappa     =     {k\Sigma}/{\gamma}$. The impedance ${\cal Z}$ is given by
    \begin{align}\label{eq:29}
    {\cal Z}(\varkappa)
    =
    -
    \frac{4 i I_eL_mL_k}{c\Sigma^2 \gamma^3I_A\sigma_{e}}
    q_1\varkappa
    e^{-\varkappa^2q_1^2/2}
    H^2
    \left(
    \varkappa
    \right)
    ,
    \end{align}
with $q_1={R_{56}^{(e,1)}\sigma_{e}\gamma}/{\Sigma}$ and $L_m$ and $L_k$ the lengths of the modulator and the kicker, respectively.
    
With one amplification section, a density perturbation with a wavenumber $k$ is amplified by the factor~\eqref{eq:26} and hence we need to multiply $\cal Z$ in Eq.~\eqref{eq:28} by $G$. Denoting the product ${\cal Z}_p ={\cal Z}G$, we find,
    \begin{align}\label{eq:30}
    {\cal Z}_p(\varkappa)
    &=
    \sqrt{2}
    A
    \frac{4 i I_eL_mL_k}{c\Sigma^2 \gamma^3I_A\sigma_{e}}
    \frac{q_1q_2\varkappa^{3/2}}{r^{1/2}}
    H^2
    \left(
    \varkappa
    \right)
    \sqrt{H(r\varkappa)}
    e^{-\varkappa^2(q_1^2+q_2^2)/2}
    \sin
    \left(
    l\sqrt{\frac{2\varkappa H(r\varkappa)}{r}}
    \right)
    ,
    \end{align}
where 
    \begin{align}\label{eq:31}
    A=
    \frac{1}{\sigma_e}
    \sqrt{\frac{I_e}{\gamma I_A}}
    ,\qquad
    l
    =
    r
    \frac{\Omega L_d}{c}
    ,\qquad
    r
    =
    \frac{\varkappa_p}{\varkappa}
    =
    \frac{\Sigma_p}{\Sigma}
    ,\qquad
    q_2
    =
    rq_p
    =
    \frac{R_{56}^{(e,2)}\sigma_{e}\gamma}{\Sigma}
    .
    \end{align}
Note the factor $r$ in the normalization of the drift $L_d$ --- since $\Omega$ scales as $1/\Sigma_p$ this extra factor $r$ makes $l$ independent of the transverse size of the electron beam in the amplification section. Replacing $\cal Z$ in Eq.~\eqref{eq:28} by ${\cal Z}_p$ we obtain
    \begin{align}\label{eq:32}
    N_\mathrm{c}^{-1}
    =
    -
    \frac{4\sqrt{2}}{\pi}
    A        
    \frac{ I_e r_h L_m L_k}{\Sigma^3 \gamma^3I_A\sigma_e\sigma_h}
    \mathrm{sign}(q_h q_1q_2)
    I_1(q_h,q_1,q_2,r,l)
    ,
    \end{align}
where
    \begin{align}\label{eq:33}
    I_1
    &=
    2 \frac{|q_h q_1q_2|}{\sqrt{r}}
    \int_{0}^\infty
    {d\varkappa}\,
    \varkappa^{5/2}
    e^{-\varkappa^2(q_1^2+q_2^2+q_h^2)/2}
    H^2
    \left(
    \varkappa
    \right)
    \sqrt{H(r\varkappa)}
    \sin
    \left(
    l\sqrt{\frac{2\varkappa H(r\varkappa)}{r}}
    \right)
    .
    \end{align}
Note the negative sign in Eq.~\eqref{eq:32}---it means that in order to have cooling one of the chicanes (or all three of them) should have a negative value of $R_{56}$ (assuming that the sin function in the integral~\eqref{eq:33} is positive). This is due to the fact that an amplification section flips the phase of the density harmonics, as was indicated at the end of the previous section. 

\begin{figure}[htb]
\centering
\includegraphics[width=0.6\textwidth, trim=0mm 0mm 0mm 0mm, clip]{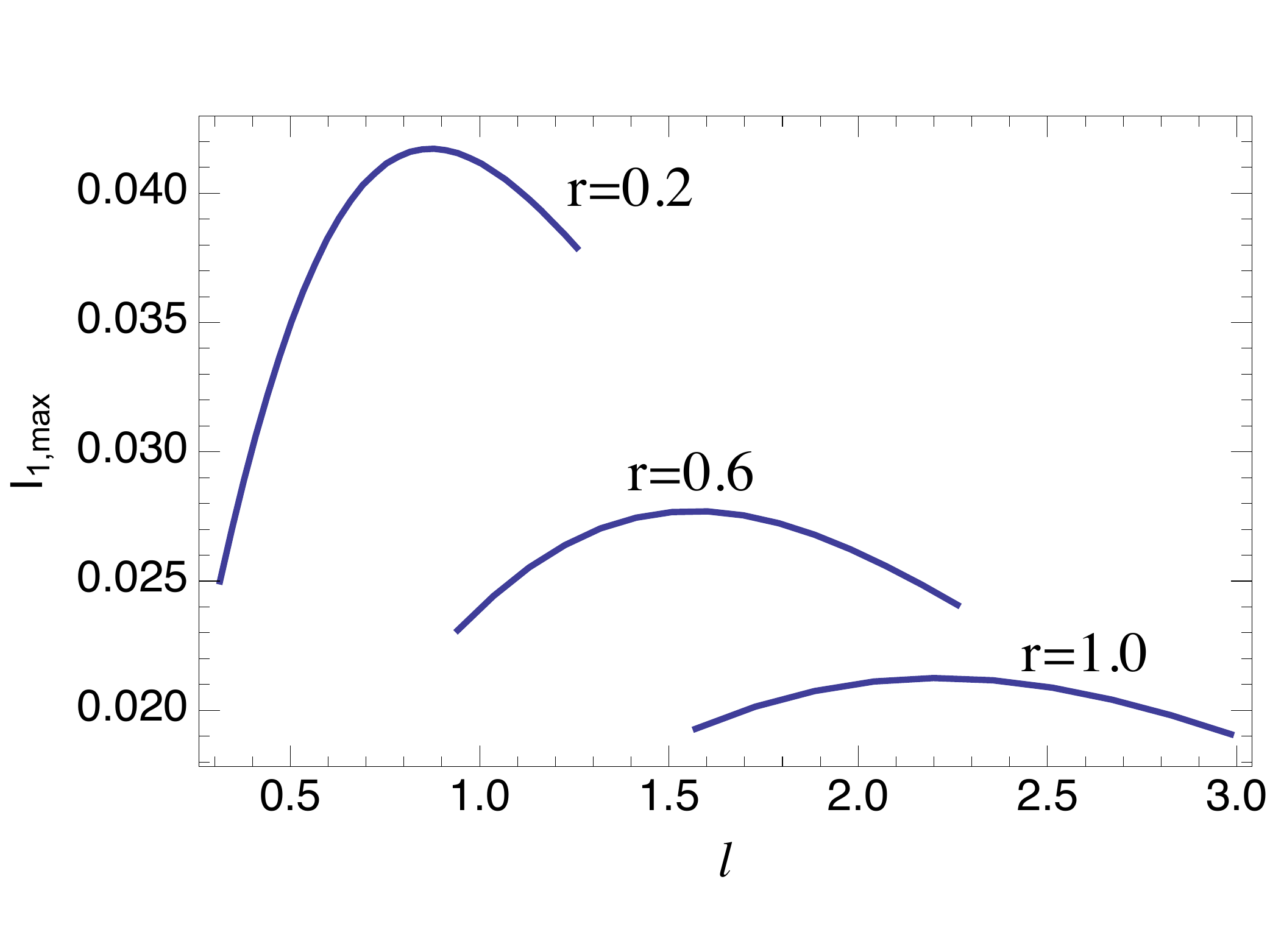}
\caption{Plot of the maximum values of $I_1$ as a function of the dimensionless length $l$ for $r=0.2$, 0.6 and 1.0.}
\label{fig:5}
\end{figure}

\begin{figure}[htb]
\centering
\includegraphics[width=0.6\textwidth, trim=0mm 0mm 0mm 0mm, clip]{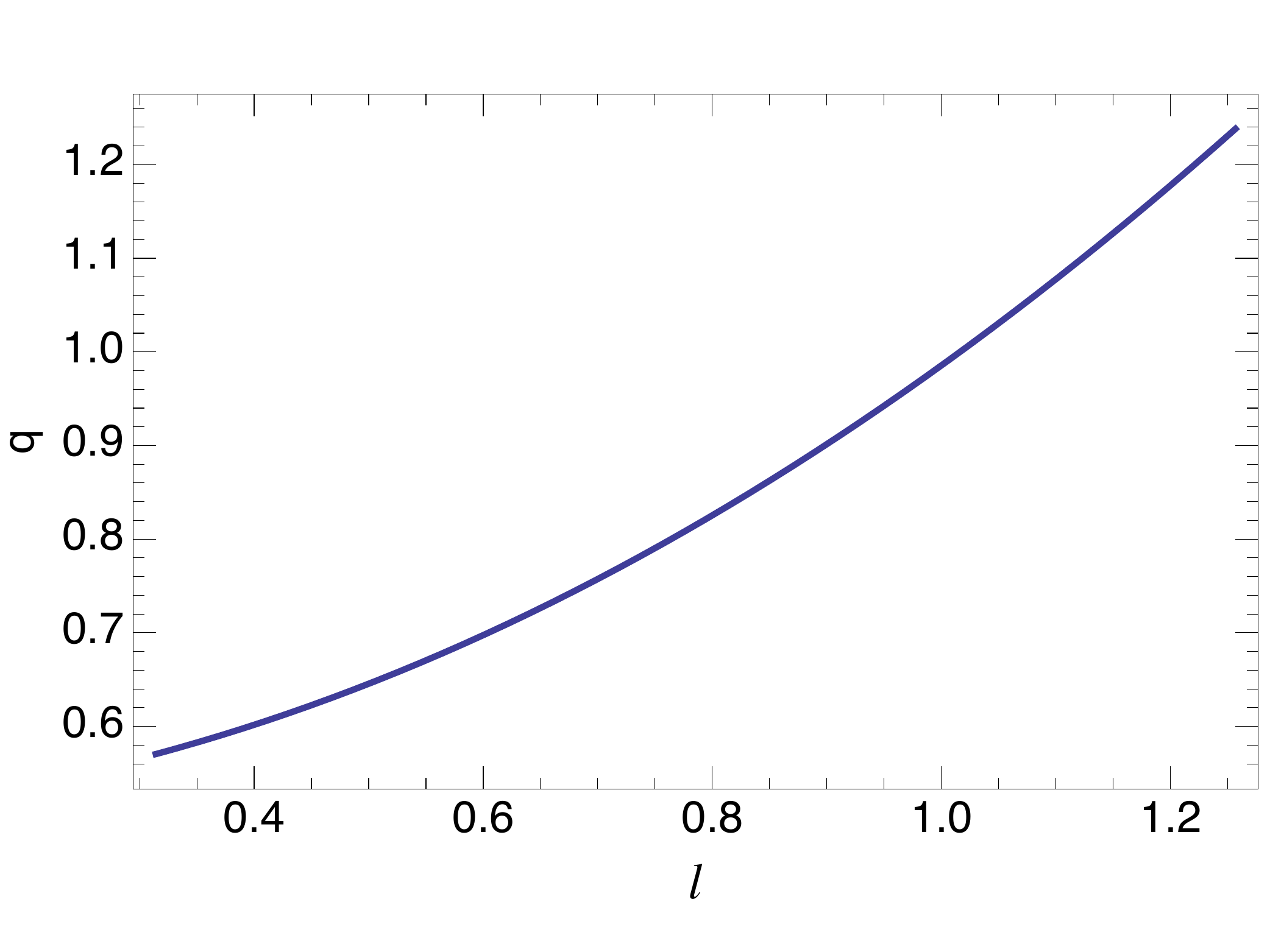}
\caption{Plot of $q$ versus $l$ for $r=0.2$.}
\label{fig:6}
\end{figure}

Because this expression is symmetric with respect to the interchange of the three variables $|q_1|$, $|q_2|$ and $|q_h|$, the maximum of $I_1$ is attained when they are all equal, $|q_h| = |q_1| = |q_2| = q$. We then have
    \begin{align}\label{eq:34}
    I_1(q,r,l)
    &=
    \frac{2 q^3}{\sqrt{r}}
    \int_{0}^\infty
    {d\varkappa}\,
    \varkappa^{5/2}
    e^{-3\varkappa^2q^2/2}
    H^2
    \left(
    \varkappa
    \right)
    \sqrt{H(r\varkappa)}
    \sin
    \left(
    l\sqrt{\frac{2\varkappa H(r\varkappa)}{r}}
    \right)
    .
    \end{align}
We numerically maximized $I_1$ with respect to $q$ for given values of the parameters $r$ and $l$. Fig.~\ref{fig:5} shows the plot of the maximum values of $I_1$ as a function of the length $l$ for three values of $r$ and Fig.~\ref{fig:6} shows, for $r=0.2$, the plot of the dimensionless chicane strength $q$ as a function of $l$ at which these maximum values of $I_1$ are attained.

We see that smaller values of $r$ (which mean a tighter focusing of the electron beam in the amplification section) allow one to obtain higher values of the integral $I_1$, and hence a higher cooling rate. Taking for the reference the value $r=0.2$, we find from Fig.~\ref{fig:5} that the maximum value of $I_{1,\mathrm{max}}$ is approximately equal to $0.042$ which gives for the cooling rate
    \begin{align}\label{eq:35}
    N_\mathrm{c}^{-1}
    =
    0.075
    \frac{I_e^{3/2} r_h L_m L_k}{\Sigma^3 \gamma^{7/2}I_A^{3/2}\sigma_e^2\sigma_h}
    .
    \end{align}
Comparing this result with Ref.~\cite{Stupakov:2018anp}, we conclude that an amplification section adds a factor of
    \begin{align}\label{eq:36}
    0.75
    \frac{1}{\sigma_e}
    \sqrt{\frac{I_e}{\gamma I_A}}
    \end{align}
to the cooling rate.

%
\section{Two amplification sections}\label{sec:6}
%

\begin{figure}[htb]
\centering
\includegraphics[width=0.6\textwidth, trim=0mm 0mm 0mm 0mm, clip]{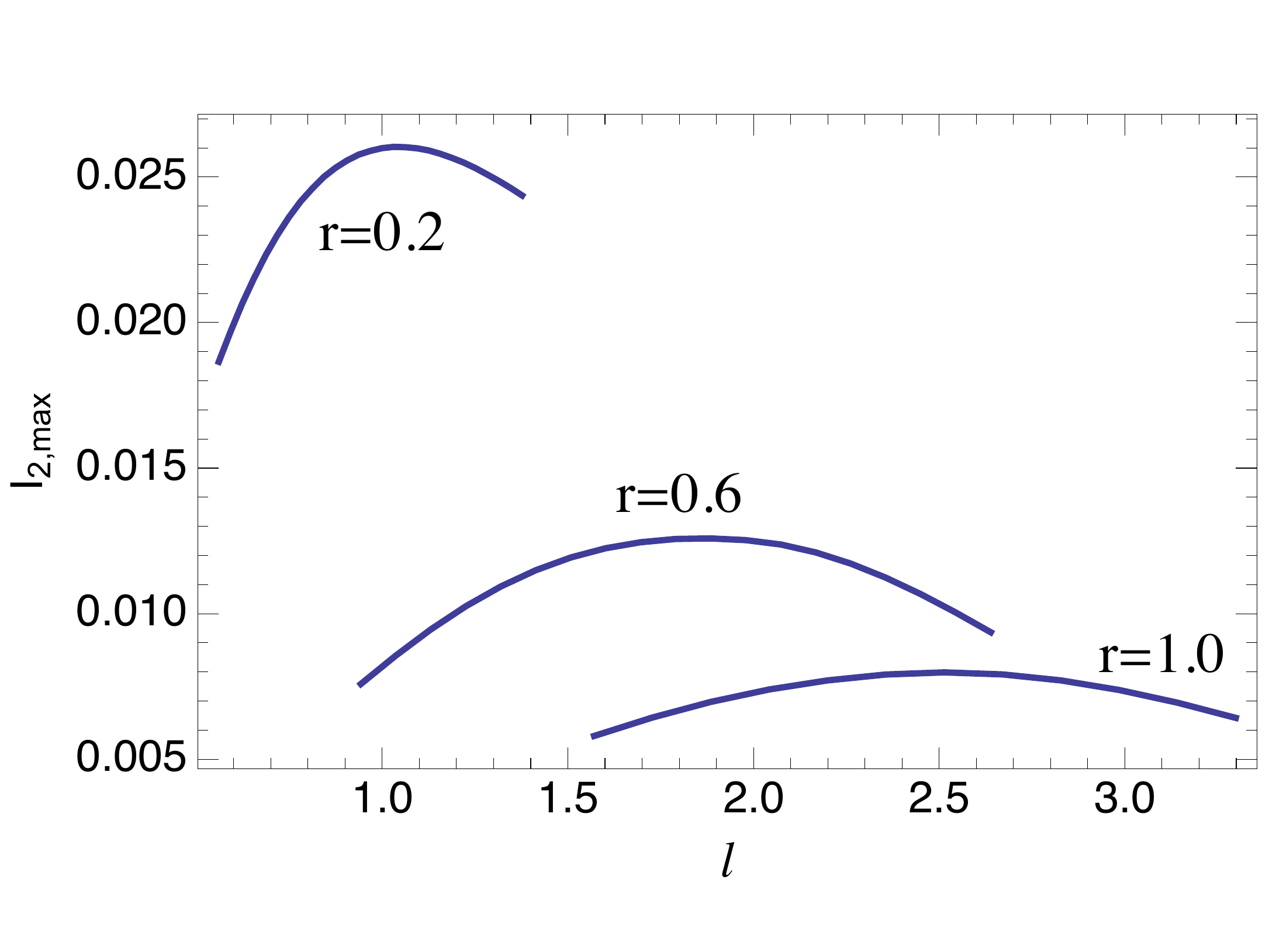}
\caption{Plot of the maximum values of $I_{2,\mathrm{max}}$ as a function of the length $l$ for $r=0.2$, 0.6 and 1.0.}
\label{fig:7}
\end{figure}

\begin{figure}[htb]
\centering
\includegraphics[width=0.6\textwidth, trim=0mm 0mm 0mm 0mm, clip]{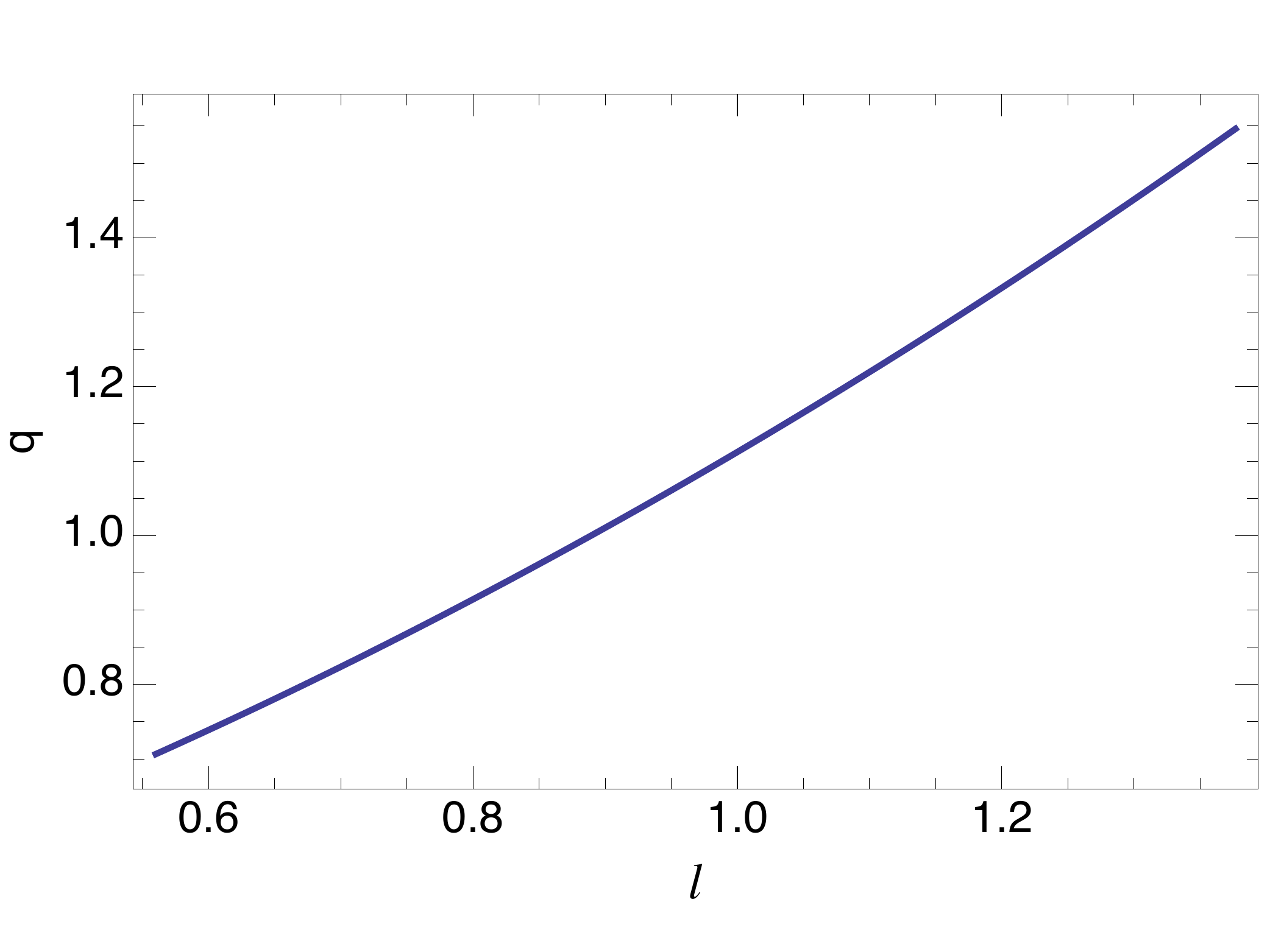}
\caption{Plot of $q$ versus $l$ for $r=0.2$.}
\label{fig:8}
\end{figure}

The setup with two amplification sections is shown in Fig.~\ref{fig:1}b. The cooling rate for a two-section amplification is obtained by replacing the impedance ${\cal Z}$ in Eq.~\eqref{eq:29} by  ${\cal Z} G^2$, which gives the following expression for the cooling rate:
    \begin{align}\label{eq:37}
    N_\mathrm{c}^{-1}
    =
    \frac{8}{\pi}
    A^2
    \frac{I_e r_h L_m L_k}{\Sigma^3 \gamma^3I_A\sigma_e\sigma_h}
    I_2(q,r,l)
    ,
    \end{align}
where
    \begin{align}\label{eq:38}
    I_2(q,r,l)
    &=
    \frac{2 q^4}{r}
    \int_{0}^\infty
    {d\varkappa}\,
    \varkappa^3
    e^{-2\varkappa^2q^2}
    H^2
    \left(
    \varkappa
    \right)
    H(r\varkappa)
    \sin^2
    \left(
    l\sqrt{\frac{2\varkappa H(r\varkappa)}{r}}
    \right)
    ,
    \end{align}
(in this expression, as in Eq.~\eqref{eq:34}, we have assumed that all dimensionless values of the chicane strength are equal, $q_h=q_1=q_2=q_3=q$). We numerically maximized $I_2$ with respect to $q$ for several values of the parameters $r$ and a range of the values of $l$, with the result shown in Fig.~\ref{fig:7}. Fig.~\ref{fig:8} shows, for $r=0.2$, the plot of the dimensionless chicane strength $q$ as a function of $l$ at which these maximum values of $I_2$ are attained.

Taking as a reference value $r=0.2$, we find from Fig.~\ref{fig:7} that the maximum value of $I_{2,\mathrm{max}}\approx 0.026$. Substituting this value into Eq.~\eqref{eq:37} we obtain for the cooling rate
    \begin{align}\label{eq:39}
    N_\mathrm{c}^{-1}
    =
    0.066
    \frac{I_e^2 r_h L_m L_k}{\Sigma^3 \gamma^4I_A^2\sigma_e^3\sigma_h}
    .
    \end{align}
Comparing this result with Ref.~\cite{Stupakov:2018anp}, we conclude that two amplification sections add a factor of $0.33 A^2$, which is approximately equal to the squared amplification factor of one section~\eqref{eq:36}.

Because the cooling rates~\eqref{eq:35} and~\eqref{eq:39} (which scale as $I_e^{3/2}$ and $I_e^2$, respectively) depend on the local electron beam current  that varies within the electron bunch, in application to practical problems, one has to average these equations taking into account the finite electron bunch length which we denote by $\sigma_{z}^{(e)}$. Assuming a Gaussian current distribution in the electron beam, $I_e = [Q_ec/\sqrt{2\pi} \sigma_{z}^{(e)}] \exp[-z^2/2 (\sigma_{z}^{(e)})^2]$, with $Q_e$ the electron bunch charge, and a Gaussian distribution for hadrons with the rms bunch length of $\sigma_{z}^{(h)}$, it is straightforward to calculate that the average values of $I_e^{3/2}$ and $I_e^2$ that a hadron sees over many passages through the electron beam:
    \begin{align}\label{eq:40}
    \langle I_e^{3/2}\rangle_z
    =
    \left(
    \frac{Q_ec}{\sqrt{2\pi}\sigma_{z}^{(e)}}
    \right)^{3/2}
    \frac{\sqrt{2}\sigma_{z}^{(e)}}{\sqrt{2(\sigma_{z}^{(e)})^2+3(\sigma_{z}^{(h)})^2}}
    ,\,\,\,
    \langle I_e^{2}\rangle_z
    =
    \left(
    \frac{Q_ec}{\sqrt{2\pi}\sigma_{z}^{(e)}}
    \right)^{2}
    \frac{\sigma_{z}^{(e)}}{\sqrt{(\sigma_{z}^{(e)})^2+2(\sigma_{z}^{(h)})^2}}
    .
    \end{align}
Here, the average is meant as an integral over the longitudinal position $z$, using the hadron probability distribution $\lambda_h(z)=(\sqrt{2\pi}\sigma_z^{(h)})^{-1}\exp(-z^2/(2(\sigma_z^{(h)})^2))$ as a weighting function. For instance, we have   
\begin{equation*}
\langle I_e^2 \rangle_z \equiv \int_{-\infty}^{\infty}dz\lambda_h(z)I_e^2(z).
\end{equation*}     
For an electron beam several times shorter than the hadron one, we can neglect $\sigma_{z}^{(e)}$ in comparison with $\sigma_{z}^{(h)}$ in Eqs.~\eqref{eq:40}. In this limit, replacing $I_e^{3/2}$ in Eq.~\eqref{eq:35} by $\langle I_e^{3/2} \rangle_z$ and $I_e^{2}$ by $\langle I_e^{2} \rangle_z$ in~\eqref{eq:39} (we recall that in these equations we have assumed the ratio $\Sigma_p/\Sigma = 0.2$) we obtain for the cooling rate with one amplification section
    \begin{align}\label{eq:41}
    N_c^{-1}
    =
    1.54\times 10^{-2}
    \frac{(Q_ec)^{3/2}r_h L_m L_k}
    {(\sigma_{z}^{(e)})^{1/2}\sigma_{z}^{(h)}\Sigma^3 \gamma^{7/2}I_A^{3/2}\sigma_e^2\sigma_h}    
    ,
    \end{align}
and with two amplification sections
    \begin{align}\label{eq:42}
    N_c^{-1}
    =
    7.4\times 10^{-3}    
    \frac{(Q_ec)^2 r_h L_m L_k}
    {\sigma_{z}^{(e)}\sigma_{z}^{(h)}\Sigma^3 \gamma^4I_A^2\sigma_e^3\sigma_h}
    .
    \end{align}
Note that in derivation of Eqs.~\eqref{eq:41} and~\eqref{eq:42} we ignored the fact that the normalized length $l$ in Eqs.~\eqref{eq:33} and~\eqref{eq:38} also depends on the local beam current (because the plasma frequency $\Omega$ scales as $\sqrt{I_e}$). This can be justified by the fact that if the argument of the $\sin$ function is chosen to be equal $\pi/2$ for the peak beam current (so that $\sin$ has a maximum value at this current), its variation near the maximum value is not so important as the direct dependence of the cooling time versus $I_e$ that is taken into account in Eqs.~\eqref{eq:41} and~\eqref{eq:42}. Nevertheless, the derived equations should be considered as a rough approximation to the cooling rates.

%
\section{Effective wake field in MBEC}\label{sec:6-1}
%

In our preceding analysis we used the concept of effective impedance ${\cal Z}_p(\varkappa)$ which is obtained from the impedance without amplification~\eqref{eq:29} by multiplying ${\cal Z}$ by the gain factor $G$ one (for one amplification section) or two (for two amplification sections) times. While ${\cal Z}_p(\varkappa)$ gives all that is needed to calculate the cooling rates, it is also instructive to analyze the wake field that is associated with this impedance. The wake field is defined by the following equation~\cite{Stupakov:2018anp}
    \begin{align}\label{eq:43}
    w(z)
    =
    -
    \frac{c}{2\pi}
    \int_{-\infty}^{\infty}
    dk\,
    {\cal Z}_p(k)
    e^{ikz}
    .
    \end{align}
It has a meaning of the energy change (normalized by the square of the hadron charge, $(Ze)^2$) induced by one hadron on a second one located at distance $z$ away due to the interaction through the electron beam (that passes through one or two amplification sections). For the case without an amplification, this wake field was calculated and plotted in Ref.~\cite{Stupakov:2018anp} where it was shown that it is positive for $z>0$ and negative for $z<0$ with a negative derivative $w'(0)>0$ at the origin.
    \begin{figure}[htb]
    \centering
    \includegraphics[width=0.6\textwidth, trim=0mm 0mm 0mm 0mm, clip]{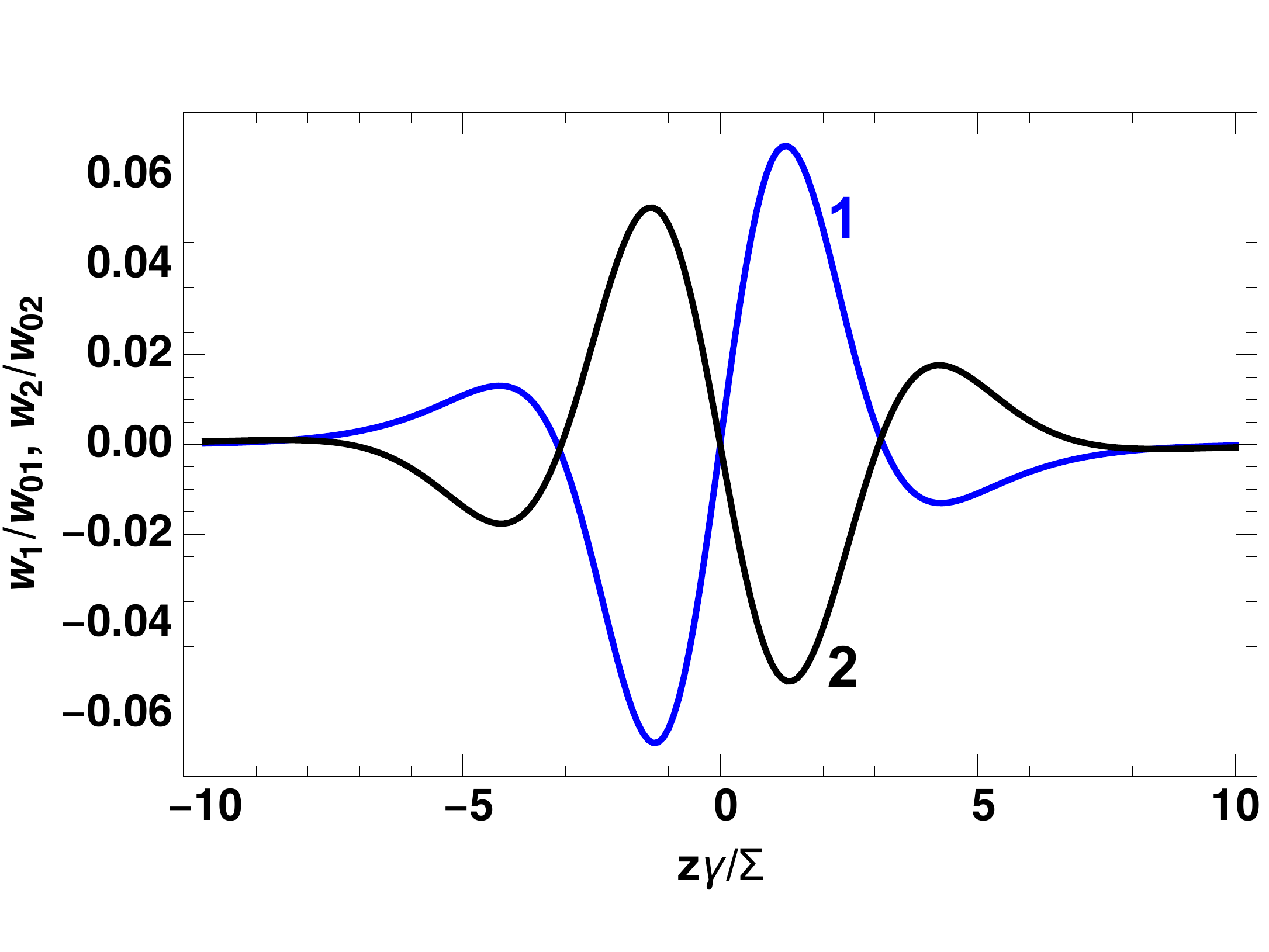}
    \caption{Plot of normalized wake functions for one (1, blue curve) and two (2, black curve) amplification sections. These wakes are odd functions of coordinate $z$.}
    \label{fig:9-1}
    \end{figure}

Plots of the wake field for one ($w_1$) and two ($w_2$) amplification sections for the optimized values $r=0.2$, $l=0.85$, and $q_e = 0.87$ are shown in Fig.~\ref{fig:9-1}. The normalization factors for these wakes are different: $w_{01} =2^{3/2} AI_eL_mL_k/\pi \Sigma^3 \gamma^2 I_A \sigma_e$ and $w_{02} = 4A^2I_eL_mL_k/\pi \Sigma^3\gamma^2I_A\sigma_e$, where $A$ is defined in Eqs.~\eqref{eq:31} and has a meaning of the amplification factor in one section (for the eRHIC parameters from Table~\ref{tab:1} the normalization factors are $ew_{01}=30.3$ V and $ew_{01}=1.1$ kV, which means that the maximum kick from a proton is $2$ V and $56$ V, respectively). Note that the wake for one amplification section (curve 1) has a positive derivative at the origin (the wake without amplification has this derivative negative) --- this corresponds to the fact mentioned in Section~\ref{sec:4} that the amplification factor of one section is negative. Also note that both wakes make oscillations and change sign away from the origin (at $z\approx \pm 3.1\Sigma/\gamma$) --- a feature absent in the wake of Ref.~\cite{Stupakov:2018anp}.  The reason of such oscillatory behavior lies in the finite bandwidth of the gain factor $G$ (see Fig.~\ref{fig:4}) which is localized in the region of small values of $\varkappa_p$. This sign change of the wake means that, for a given value of $R_{56}^{(h)}$, hadrons with a large energy deviation will be shifted longitudinally into the region where the cooling force changes sign and leads to further increase of the relative energy (the so called anti-cooling effect).   Effects of these nature have been studied for classical stochastic cooling (see, e.g., \cite{Lebedev_2016_cooling}) and for the optical stochastic cooling \cite{PhysRevSTAB.15.032801}; they impose a constrain on the value of $R_{56}^{(h)}$ in order to avoid the anti-cooling for particles at the tail of the distribution function.

%
\section{Computer simulations}\label{sec:7}
%

To test our analytical theory we carried out computer simulations of MBEC with one amplification section. In these simulations, electrons and hadrons are represented by macroparticles that interact with the force given by Eq.~\eqref{eq:1}. Initially, $N_e$ electron macroparticles  are randomly distributed  in the interval $0<z<\Delta z$ with the energy $\eta^{(e)}_i$ of the $i$-th electron randomly assigned from a Gaussian distribution with the rms width $\sigma_e$. Periodic boundary conditions are set at the boundaries of the interval $[0,\Delta z]$. A hadron particle, with an energy $\eta^{(h)}$ randomly selected from a Gaussian distribution with the rms width $\sigma_h$, is placed at a random location within the interval and the energy of each electron $i$ is changed by $\Delta\eta_i^{(e)} = f_{z,i}L_m/\gamma m_ec^2$, where $f_{z,i}$ is the force exerting by the hadron on electron $i$. On the next step, corresponding to the passage of the hadrons through the chicane, the hadron is shifted longitudinally by $R_h\eta^{(h)}$. The electrons pass through the chicane $R_{56}^{(e,1)}$ where they are shifted longitudinally by $R_{56}^{(e,1)}(\eta^{(e)}_i+\Delta\eta^{(e)}_i)$ and then through one or two amplification sections, as shown in Fig~\ref{fig:1}. Finally, in the kicker, the hadron energy is changed from $\eta^{(h)}$ to $\eta^{(h)}+\Delta\eta^{(h)}$ with $\Delta\eta^{(h)}=\sum_{i=1}^{N_e} f_{z,i}L_k/\gamma m_hc^2$, where now $f_{z,i}$ denotes the force acting on the hadron from $i$th electron. This procedure is repeated $M$ times and the cooling rate is estimated  as an average over $M$ runs of the difference $(\eta^{(h)}+\Delta\eta^{(h)})^2-\sigma_h^2$.

In the drift sections of the amplification cascades we use the following equations of motion for the electrons,
    \begin{align}\label{eq:44}
    \frac{d\eta_i}{cdt}
    &=
    \frac{r_e}{\gamma\Sigma_p^2}
    \sum_{i\ne j}
    \Phi
    \left(
    \gamma\frac{z_i-z_j}{\Sigma_p}
    \right)
    ,
    \nonumber\\
    \frac{d z_i}{cdt}
    &=
    \frac{\eta_i}{\gamma^2}
    ,
    \end{align}
where $\Sigma_p$ is the rms size of the beam in the drift.
We scale the energy deviation, $\tilde p = \eta\sqrt{\Sigma_p/r_e}$, the coordinate $\tilde z = z\gamma/\Sigma_p$, and the distance, $\tilde s = ct/\tilde l$, with $\tilde l=\gamma (\Sigma_p^3 /r_e)^{1/2}$, so that the equations of motion become dimensionless, 
    \begin{align}\label{eq:45}
    \frac{d\tilde p_i}{d\tilde s}
    &=
    \sum_{i\ne j}
    \Phi
    \left(
    \tilde z_i-\tilde z_j
    \right)
    ,
    \nonumber\\
    \frac{d \tilde z_i}{d\tilde s}
    &=
    \tilde p_i
    .
    \end{align}
The plasma frequency for the wavelengths with $\kappa\gg 1$ is given by $\Omega$ in Eq.~\eqref{eq:18} which means that the plasma period in variable $\tilde s$ is
    \begin{align}\label{eq:46}
    \Delta \tilde s_p
    =
    \frac{2\pi}{\sqrt{\nu}}
    ,
    \end{align}
where $\nu = n_0\Sigma_p/\gamma$.

A different normalization of the variables is used in the modulator and the kicker: the energy deviation $\eta$ is normalized by the rms energy spread of the electron beam, $q=\eta/\sigma_e$, and $z$ is normalized by the transverse size $\Sigma$ of the beam in the modulator, $\zeta = z\gamma/\Sigma$. With this normalization the energy change of an electron on length $L_m$ due to an interaction with a hadron is
    \begin{align}\label{eq:47}
    \Delta q_i
    =
    -
    \frac{Zr_eL_m}{\gamma\Sigma^2\sigma_e}
    \sum_{i\ne j}
    \Phi
    \left(
    \zeta_i-\zeta_j
    \right)
    .
    \end{align}
We denote by $A_1$ the factor in front of the sum in this equation. Note the relations between the variables $\tilde z$ and $\tilde p$ in the amplifier and $\zeta$ and $q$ in the kicker and the modulator:
    \begin{align}\label{eq:48}
    \tilde z_i
    =
    \zeta_i 
    \frac{\Sigma}{\Sigma_p}
    ,\qquad
    \tilde p_i
    =
    q_i\sigma_e
    \sqrt{\frac{\Sigma_p}{r_e}}
    .
    \end{align}

We first simulated the amplification of initial perturbations of small amplitude in the electron beam, as discussed in Section~\ref{sec:4}. We used $N_e=10^5$ electron macroparticles and the length of the ``electron bunch'' $\Delta z = 20\Sigma_p/\gamma$ in the simulations. An initial density perturbation with the dimensionless wavenumber $\varkappa_p$ and a relative amplitude of the density perturbation of $10^{-3}$ was launched and propagated over the distance of one-quarter of the plasma oscillation (for that value of $\varkappa_p)$ and then sent through a chicane. The strength of the chicane was optimized to obtain the maximum amplitude of the density modulation at the exit. The simulated amplification factor $G_\mathrm{max}$, after averaging over $M=100$ runs and scaling by the parameter $-{\sigma_e}^{-1} \sqrt{{I_e}/{\gamma I_A}}$, is plotted by red symbols in Fig.~\ref{fig:4}. One can see an excellent agreement with the theoretical formula~\eqref{eq:27}.

\begin{figure}[htb!]
    \centering
    \includegraphics[width=0.6\textwidth, trim=0mm 0mm 0mm 0mm, clip]{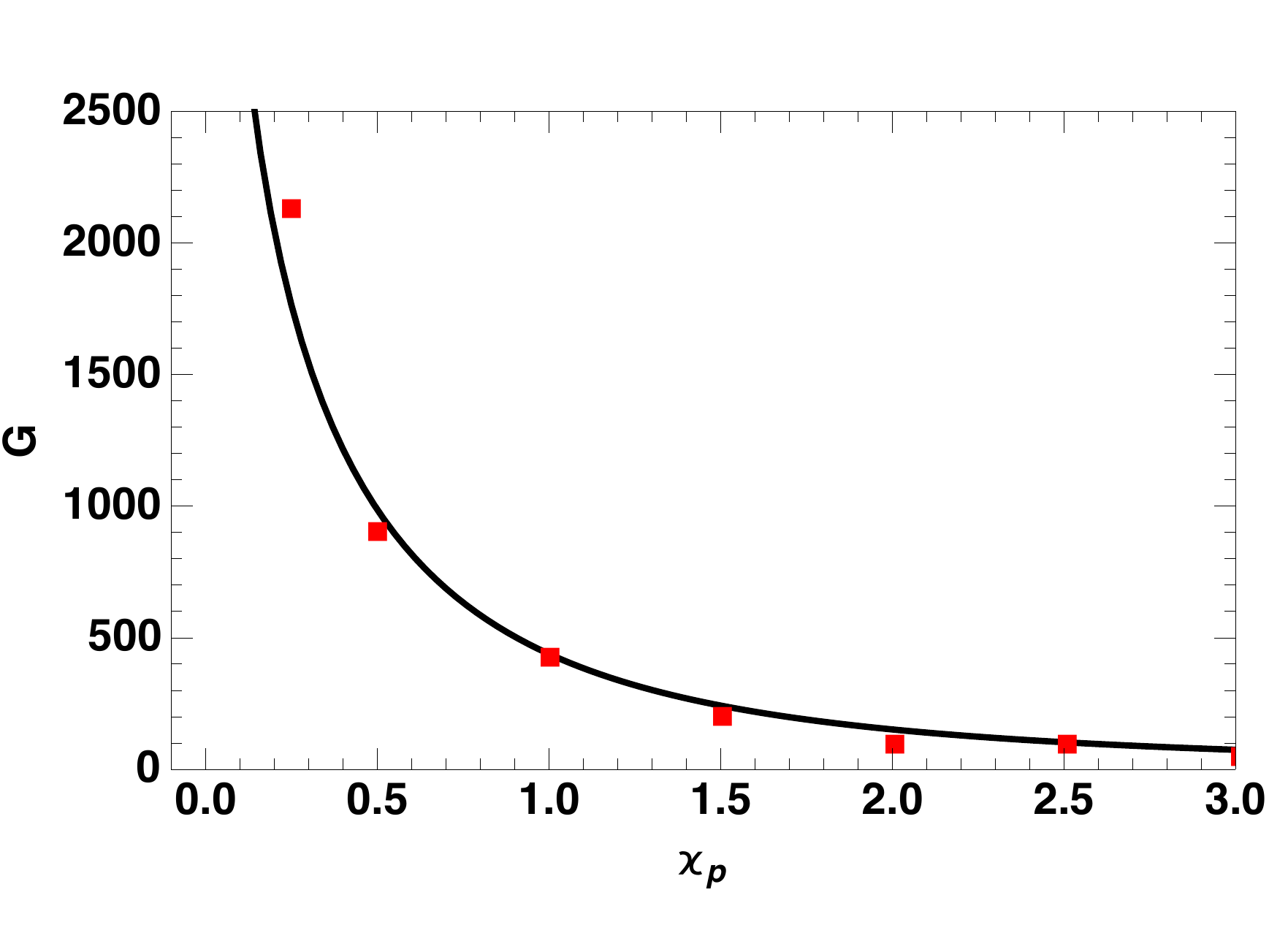}
    \caption{Amplification factor of a small initial perturbation for two sections.}
    \label{fig:9}
\end{figure}

We also simulated the amplification factor of an initial small perturbation in two amplification sections. In this case we used $N_e=10^5$ macroparticles and the simulation interval $\Delta z = 50\Sigma_p/\gamma$ with averaging over $M=100$ runs. The initial amplitude of the density perturbation was $10^{-4}$ and the parameter $A$ was chosen to be equal to 1 (see Eq.~\eqref{eq:31} for the definition of $A$). The result is shown in Fig.~\ref{fig:9}, where the red symbols are
the simulated gain $G$ and the solid curve is the square of the one-section amplification Eq.~\eqref{eq:27} with the sin function replaced by unity. The scatter of the points in this figure are due to the amplification of the intrinsic noise in the electron beam.

By properly scaling all dimensional variables of the simulation problem, one can find that it is determined by several dimensionless parameters. The first one, $\nu = n_{0e}\Sigma/\gamma$, is equal to the number of electrons on the length $\Sigma/\gamma$ and is proportional to the electron beam current. Two more parameters, $A_1$ and $A_2$, characterize the interaction strength in the modulator and the kicker normalized by the electron and hadron energy spread, respectively,
    \begin{align}\label{eq:49}
    A_1
    =
    \frac{Zr_eL_m}{\gamma\Sigma^2\sigma_e}
    ,\qquad
    A_2
    =
    \frac{r_hL_k}{Z\gamma\Sigma^2\sigma_h}
    ,
    \end{align}
and parameter $A$ from Eq.~\eqref{eq:31} is related to the amplification factor of one cascade. Two more parameters are the dimensionless strengths of the chicanes, $q_e$ and $q_h$, defined in Sections~\ref{sec:6} and~\ref{sec:7}. Finally, there is a ratio $r=\Sigma_p/\Sigma$ of the transverse size of the electron beam in the amplification section and in the modulator and the kicker. 

Calculating numerical values of $\nu$, $A$, $A_1$ and $A_2$ for the eRHIC parameters from Table~\ref{tab:1} and assuming the electron peak current of $I_e = 30$ A, we find
    \begin{align}\label{eq:50}
    \nu = 1.5\times 10^6
    ,\qquad
    A   = 24.5
    ,\qquad
    A_1 = 7.8\times 10^{-6}
    ,\qquad
    A_2 = 9.3\times 10^{-10}
    .
    \end{align} 
Simulations with these values are difficult due to a required large number of macroparticles and small values of the interaction strengths, so we used larger values for $A_1$ and $A_2$ and smaller values for $\nu$ and $A$:
    \begin{align}\label{eq:51}
    \nu = 5\times 10^4
    ,\qquad
    A   = 10
    ,\qquad
    A_1    = 1\times 10^{-3}
    ,\qquad
    A_2    = 1\times 10^{-4}
    .
    \end{align} 
Because $A_1$ and $A_2$ are proportional to the square of the charge, the increased values of $A_2$ and $A_1$ can be interpreted as if macroparticles carry a charge larger than the elementary charge $e$. Our parameter choice~\eqref{eq:51} can be interpreted as if each electron macroparticle has a charge of approximately $11 e$ and each hadron has a charge $328 e$ (assuming $Z=1$).

\begin{figure}[htb]
    \centering
    \includegraphics[width=0.6\textwidth, trim=0mm 0mm 0mm 0mm, clip]{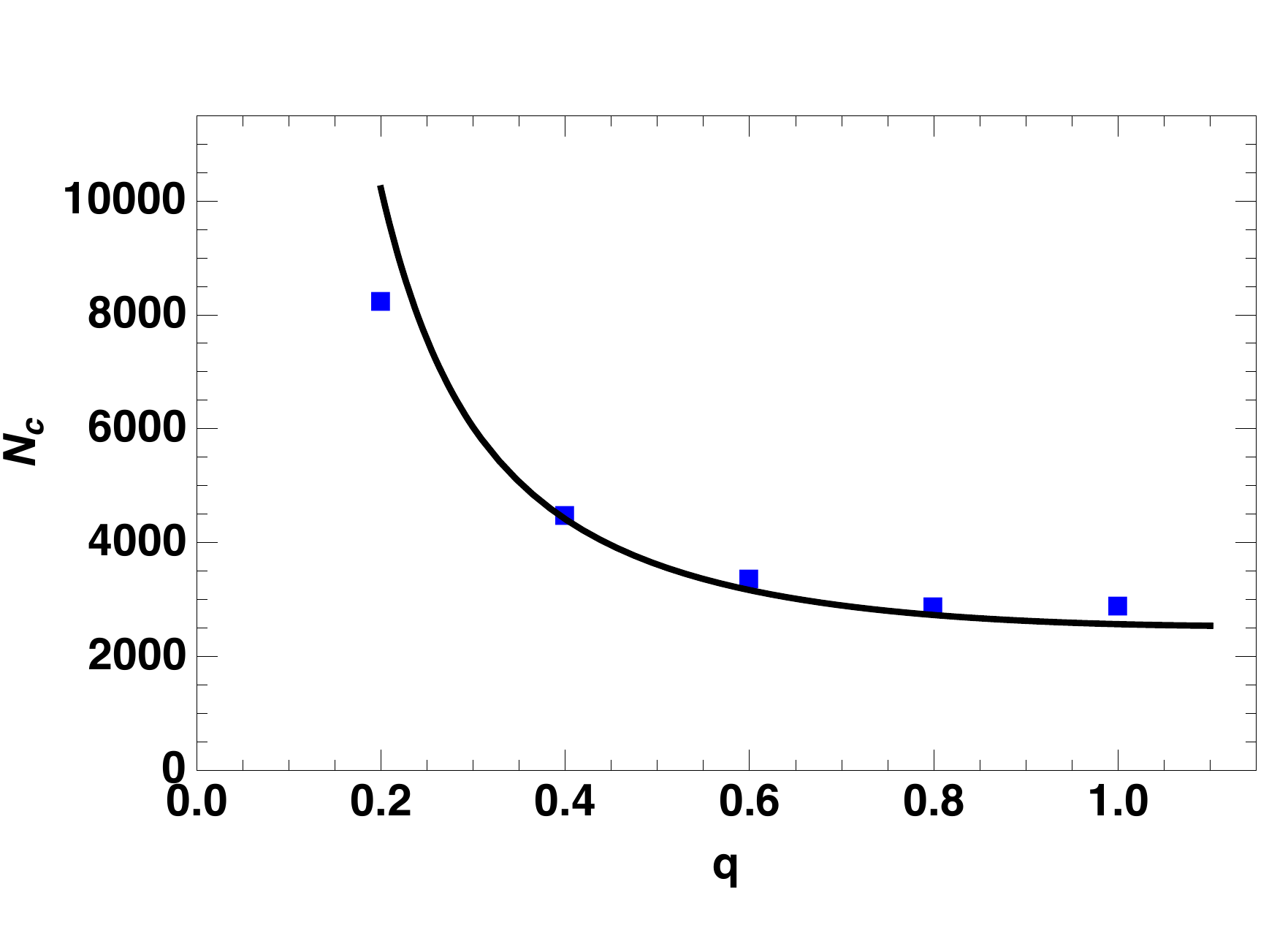}
    \caption{Cooling time as a function of dimensionless chicane strength for a system with one amplification section.}
    \label{fig:10}
    \end{figure}

With the dimensionless simulation parameters given by Eq.~\eqref{eq:51} we also simulated the cooling process with one amplification section. In this simulation, we used $N_e=10^5$ electron macroparticle and the length of the ``electron bunch'' $\Delta z = 10 \Sigma/\gamma$ in the simulations. The averaging was done over $M=5\times10^4$ runs. The plot of the simulated cooling times as a function of the dimensionless chicane strength $q$ is shown in Fig.~\ref{fig:10} by blue squares.
The solid curve is calculated using Eq.~\eqref{eq:32} with $q_h=q_1=p_2=q$. One can see that Eq.~\eqref{eq:32} is in good agreement with the simulations which we consider as a confirmation of the correctness on our analytical results. In these simulations we assumed the ratio $\Sigma_p/\Sigma = 1$.

\begin{figure}[htb]
    \centering
    \includegraphics[width=0.6\textwidth, trim=0mm 0mm 0mm 0mm, clip]{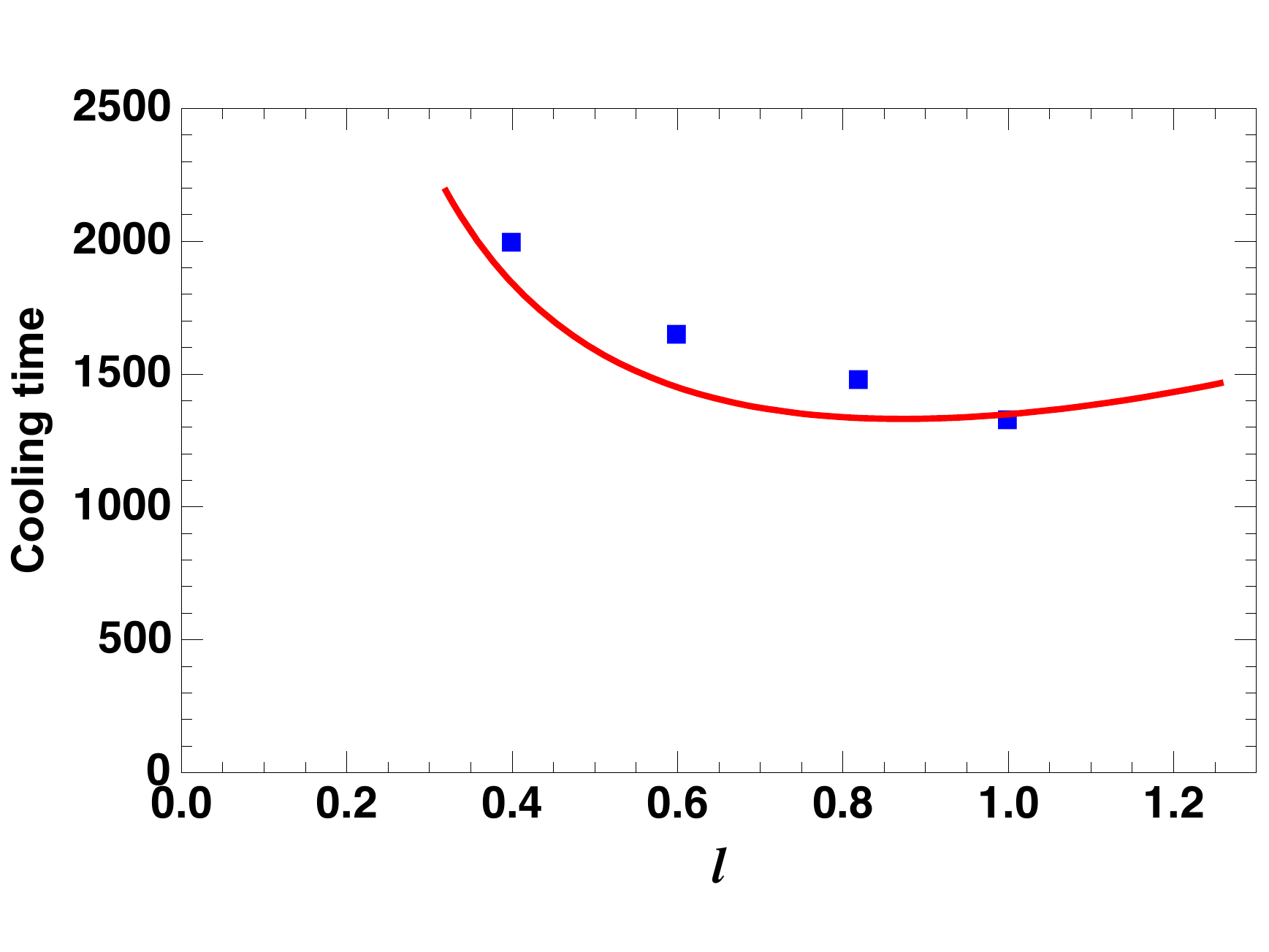}
    \caption{Cooling time as a function of dimensionless length $l$ of the amplification section for one cooling section.}
    \label{fig:11}
    \end{figure}

In another set of simulations we used the ratio $\Sigma_p/\Sigma = 0.2$ and varied the length $l$ of the amplification section. The result of these simulations is shown in Fig.~\ref{fig:11}. The theoretical curve is calculated with the same Eq.~\eqref{eq:32}.
Again, we find a reasonably good agreement between theory and the simulation.

%
\section{Saturation and noise effects}\label{sec:8}
%

In this section, we discuss the effects of diffusion and nonlinear behavior (saturation) for an MBEC configuration that utilizes amplification stages. In Ref.~\cite{Stupakov:2018anp}, it was shown that diffusion effects due to the noise in the hadron beam can be quantified by means of a diffusion coefficient $D_h$, which is given by
\begin{align}\label{eq:52}
D_h
=
\frac{n_{0h}}{4\pi T}
\left(
\frac{r_hc}{ \gamma}
\right)^2
\int_{-\infty}^\infty
{dk}
|{\cal Z}(k)|^2
.
\end{align}
Here, $\cal Z$ is the impedance without amplification cascades, given by~\eqref{eq:29}. For the cooling to overcome the diffusion, we need to satisfy the following requirement:
\begin{align}\label{eq:53}
D_h<\frac{\sigma_h^2}{2t_c}=\frac{\sigma_h^2}{2TN_c}\,.
\end{align}
With one stage of amplification, we need to replace $\cal Z$ with ${\cal Z}_p={\cal Z}G$, where $G$ is the amplification gain of Eq.~\eqref{eq:26}. For the case of two amplification stages, we instead replace $\cal Z$ by ${\cal Z} G^2$. Here, we will focus on the latter case, for which the effects of diffusion and saturation are more important. After working out the algebra, we obtain
\begin{equation}\label{eq:54}
D_h=
\frac{16\sigma_h^2}{\pi T}
\frac{I_hI_e^4r_h^2L_m^2L_k^2}{I_A^5\gamma^9r_e\Sigma^5\sigma_e^6\sigma_h^2}\frac{q^6}{r^2}
\int_{-\infty}^\infty
{d\varkappa}
\varkappa^4
H^4(\varkappa)
H^2(r\varkappa)
e^{-3\varkappa^2q^2}
\sin^4
\left(
l\sqrt{\frac{2\varkappa H(r\varkappa)}{r}}
\right)
.
\end{equation}
The equation given above assumes that all chicane strengths, $q$, are equal to each other which is indeed the optimized configuration. To take the finite length of the electron beam into consideration, we follow an averaging procedure entirely analogous to the one used for the cooling. In particular, after noting that $D_h\propto I_hI_e^4$, we obtain
\begin{align}\label{eq:55}
&\langle D_h\rangle_z\approx
\frac{4\sigma_h^2}{\pi^3 T}
\frac{(Q_ec)^4I_h^{(0)}r_h^2L_m^2L_k^2}{I_A^5(\sigma_z^{(e)})^3\sigma_z^{(h)}\gamma^9r_e\Sigma^5\sigma_e^6\sigma_h^2}\nonumber\\
&\times
\frac{q^6}{r^2}
\int_{-\infty}^\infty
{d\varkappa}
\varkappa^4
H^4(\varkappa)
H^2(r\varkappa)
e^{-3\varkappa^2q^2}
\sin^4
\left(
l\sqrt{\frac{2\varkappa H(r\varkappa)}{r}}
\right)
,
\end{align}
where we have neglected the dependence of $l$ with respect to $I_e$. Moreover, we have made use of the property
\begin{align}\label{eq:56}
\langle I_e^{m}I_h^{n}\rangle_z
=
\left(\frac{Q_ec}{\sqrt{2\pi}\sigma_{z}^{(e)}}
\right)^{m}
\frac{(I_h^{(0)})^n\sigma_{z}^{(e)}}{\sqrt{(n+1)(\sigma_{z}^{(e)})^2+m(\sigma_{z}^{(h)})^2}}
\approx
\left(\frac{Q_ec}{\sqrt{2\pi}\sigma_{z}^{(e)}}
\right)^{m}
\frac{(I_h^{(0)})^n\sigma_{z}^{(e)}}{\sqrt{m}\sigma_{z}^{(h)}}
,
\end{align}
which generalizes Eq.~\eqref{eq:40}. For $r=0.2$, $l=1.0$ and $q=1.1$ (the optimized section length and chicane strength), the value of the term in the second line of ~\eqref{eq:55} is about $10^{-3}$. Combining this with Eq.~\eqref{eq:42}, we obtain the formula
\begin{equation}\label{eq:57}
r_1\equiv\frac{2\langle D_h\rangle_z}{(\sigma_h^2/T)\langle 1/N_c\rangle_z} 
\approx 0.019\frac{(Q_ec)^2I_h^{(0)}r_hL_mL_k}{I_A^3(\sigma_z^{(e)})^2\gamma^5r_e\Sigma^2\sigma_e^3\sigma_h}
.
\end{equation}

Using a similar procedure, we can derive the diffusion rate of the hadrons due to the intrinsic noise in the electron beam. For the energy perturbation of the hadrons due to the electrons in the kicker, we use a formula analogous to Eq.~(48) from Ref.~\cite{Stupakov:2018anp} i.e.
\begin{align}\label{eq:58}
\Delta\eta^{(h)}(z)
=
\int_{-\infty}^\infty
dz'
\delta n_e(z')
G_\eta^{(h)}(z-z')
,
\end{align}
with
\begin{align}\label{eq:59}
G_\eta^{(h)}(z)
=
-
\frac{r_hL_k}{Z\gamma \Sigma^2}
\Phi
\left(\frac{z\gamma}{\Sigma}\right)
.
\end{align}
The electron density perturbation $\delta n_e$ is now due to the shot noise in the e-beam, so we have
\begin{align}\label{eq:60}
\langle \delta n_e(z) \delta n_e(z') \rangle = n_{0e}\delta(z-z')
\end{align}
for the case of no amplification. A general definition of the diffusion coefficient was given in Ref.~\cite{Stupakov:2018anp} as
\begin{align}\label{eq:61}
D
=
\frac{1}{2T}
\langle
(\Delta \eta^{(h)})^{2}
\rangle
.
\end{align}
Adapted for the new coefficient, this expression yields
\begin{align}\label{eq:62}
D_e
&=
\frac{1}{2T}
\int_{-\infty}^\infty
dz'\,dz''\
G_\eta^{(h)}(z-z')
G_\eta^{(h)}(z-z'')
\langle
\delta n_e(z')
\delta n_e(z'')
\rangle
\nonumber\\
&=
\frac{1}{2T}
n_{0e}
\int_{-\infty}^\infty
dz\,
G_\eta^{(h)}(z)^2
=
n_{0e}
\frac{r_h^2L_k^2}{TZ^2\gamma^3 \Sigma^3}
\int_{0}^\infty
d\xi\,
\Phi(\xi)^2
.
\end{align}
To accommodate the amplification effect, we first substitute 
\begin{align}\label{eq:63}
\int_{0}^\infty
d\xi\,
\Phi(\xi)^2
=
\frac{2}{\pi}
\int_{0}^\infty
d\varkappa\,
H(\varkappa)^2
\end{align}
into Eq.~\eqref{eq:62}, to obtain
\begin{align}\label{eq:64}
D_e
&=
\frac{2}{\pi}
n_{0e}
\frac{r_h^2L_k^2}{TZ^2\gamma^3 \Sigma^3}
\int_{0}^\infty
d\varkappa\,
H(\varkappa)^2
.
\end{align}
Next, we note that, when including amplification stages, the Fourier quantity $\tilde G_k\propto\int_{-\infty}^{\infty}dze^{-ikz}G_\eta^{(h)}\propto H(\varkappa)$ is modified by a factor of $G^S$, where $S$ is the number of stages and $G$ is the gain factor expressed by Eq.~\eqref{eq:26}. Thus, since $H(\varkappa)$ is basically akin to the impedance ${\cal Z}$ in Eq.~\eqref{eq:52}, Eq.~\eqref{eq:64} becomes
\begin{align}\label{eq:65}
D_e
&=
\frac{2}{\pi}
n_{0e}
\frac{r_h^2L_k^2}{TZ^2\gamma^3 \Sigma^3}
\int_{0}^\infty
d\varkappa\,
H(\varkappa)^2|G(\varkappa)|^4
,
\end{align}
for the case of two amplification stages. Substituting the expression for the gain, we obtain
\begin{equation}\label{eq:66}
D_e=
\frac{8\sigma_h^2}{\pi T}
\frac{I_e^3r_h^2L_k^2}{Z^2I_A^3\gamma^5r_e\Sigma^3\sigma_e^4\sigma_h^2}\frac{q^4}{r^2}
\int_{0}^\infty
{d\varkappa}
\varkappa^2
H^2(\varkappa)
H^2(r\varkappa)
e^{-2\varkappa^2q^2}
\sin^4
\left(
l\sqrt{\frac{2\varkappa H(r\varkappa)}{r}}
\right)
.
\end{equation}
The averaged diffusion rate is 
\begin{align}\label{eq:67}
&\langle D_e\rangle_z\approx
\frac{8\sigma_h^2}{\sqrt{3}\pi(2\pi)^{3/2} T}
\frac{(Q_ec)^3r_h^2L_k^2}{Z^2I_A^3(\sigma_z^{(e)})^2\sigma_z^{(h)}\gamma^5r_e\Sigma^3\sigma_e^4\sigma_h^2}\nonumber\\
&\times
\frac{q^4}{r^2}
\int_{0}^\infty
{d\varkappa}
\varkappa^2
H^2(\varkappa)
H^2(r\varkappa)
e^{-2\varkappa^2q^2}
\sin^4
\left(
l\sqrt{\frac{2\varkappa H(r\varkappa)}{r}}
\right)
,
\end{align}
yielding a new diffusion ratio (again for $r=0.2$, $l=1.0$ and $q=1.1$)
\begin{equation}\label{eq:68}
r_2\equiv\frac{2\langle D_e\rangle_z}{(\sigma_h^2/T)\langle 1/N_c\rangle_z} 
\approx 
0.45\frac{Q_ecr_h(L_k/L_m)}{Z^2I_A\sigma_z^{(e)}\gamma r_e\sigma_e\sigma_h}
.
\end{equation}

Finally, we would like to address the issue of possible nonlinear behavior in the amplification cascade, an effect which can be important if the gain is large enough. Recalling our earlier analysis, we observe that the linearization of the Vlasov equation is valid provided that $|\delta n|\ll n_{0e}$, where $\delta n=\int_{-\infty}^{\infty}dk\,e^{ikz}\delta\hat n_k/2\pi$ is the density modulation of the electron beam. Let us assume, for the moment, that no amplification stages are present and $\delta n$ represents the density perturbation after the electron chicane. In Ref.~\cite{Stupakov:2018anp} it was shown that $\delta\hat n_k={\cal F}(k)\delta\hat n_k^{(M)}$, where $\delta\hat n_k^{(M)}$ is the density perturbation of the hadrons in the modulator and 
${\cal F}(k)\equiv Zn_{0e}g_0(k)\zeta_0(k)$. The other functions mentioned here are given by

\begin{align}\label{eq:69}
\zeta_0(k)
=-\frac{2ir_eL_m}{\Sigma\gamma^2}
H\left(\frac{k\Sigma}{\gamma}\right)
,
\end{align}
and
\begin{align}\label{eq:70}
g_0(k)
=ikR_{56}^{(e,1)}
e^{-k^2(R_{56}^{(e,1)})^2\sigma_{e}^2/2}
.
\end{align}
These are basically the expressions of Eqs.~(53), (55) and (57) from Ref.~\cite{Stupakov:2018anp}, with some minor notation changes. In the case of amplification stages, we have ${\cal F}(k)\equiv Zn_{0e}g_0(k)\zeta_0(k)G^S(k)$, where (again) $S$ is the number of stages and $G$ is the gain factor. In all these cases, we have ${\cal F}(-k)={\cal F}^{*}(k)$ so the linearity condition can be re-written as
\begin{align}\label{eq:71}
I_{\rm sat}^2 &\equiv \langle\delta n^2\rangle/n_{0e}^2 = \frac{1}{(2\pi)^2n_{0e}^2}\int_{-\infty}^{\infty}dkdk'e^{i(k+k')z}{\cal F}(k){\cal F}(k')\delta\hat n_k^{(M)}\delta\hat n_{k'}^{(M)}\nonumber\\
& = \frac{n_{0h}}{2\pi n_{0e}^2}\int_{-\infty}^{\infty}dk|{\cal F}(k)|^2\ll 1\,,
\end{align}
where we have defined a saturation measure $I_{\rm sat}$ and made use of the property $\langle \delta\hat n_k^{(M)}\delta\hat n_{k'}^{(M)}\rangle=2\pi n_{0h}\delta(k+k')$ regarding the initial noise in the hadron beam. Collecting all the necessary terms, the result for two stages becomes
\begin{equation}\label{eq:72}
I_{\rm sat}^2 = \frac{16Z^2r_eL_m^2I_hI_e^2}{\pi\gamma^5\Sigma^3\sigma_e^6I_A^3}
\frac{q^6}{r^2}
\int_{0}^\infty
{d\varkappa}
\varkappa^4
H^2(\varkappa)
H^2(r\varkappa)
e^{-3\varkappa^2q^2}
\sin^4\left(l\sqrt{\frac{2\varkappa H(r\varkappa)}{r}}\right)\ll 1
.
\end{equation}

%
\section{Estimates for the {e}RHIC collider}\label{sec:9}
%

As a numerical illustration of the general theory developed in the previous sections we will estimate the optimized cooling rate for the nominal parameters of the electron-hadron collider eRHIC~\cite{Montag:IPAC2017}. The parameters of the proton beam in eRHIC and hypothetical parameters of the electron beam in the cooling system are given in  Table~\ref{tab:1}.
\begin{table}[hbt]
    \begin{center}
        \begin{tabular}{lc}
            \hline
            \hline
            Proton beam energy\hspace{70mm}        &  275 GeV \\
            RMS length of the proton beam, $\sigma_{z}^{(h)}$         &  5 cm \\
            RMS relative energy spread of the proton beam, $\sigma_{h}$& $4.6\times 10^{-4}$\\
            Peak proton beam current, $I_h$ & 23 A\\
            Peak electron beam current, $I_e$ & 30 A\\
            RMS transverse size of the beam in the cooling section, $\Sigma$ & 0.7 mm\\
            Electron beam charge, $Q_e$ & 1 nC\\
            RMS relative energy spread of the electron beam, $\sigma_{e}$& $1\times 10^{-4}$\\
            Modulator and kicker length, $L_m$ and $L_k$  & 40 m\\
            \hline
            \hline
        \end{tabular}
        \caption{Parameters of the eRHIC collider with a hypothetical MBEC cooling section.}
        \label{tab:1}
    \end{center}
\end{table}

Substituting parameters from Table~\ref{tab:1} into Eqs.~\eqref{eq:41} and \eqref{eq:42} gives    $ N_\mathrm{c} = 7.7\times 10^{8}$ for one-section amplification and $N_\mathrm{c} = 4.1\times 10^{7}$ for two sections. With the revolution period in the RHIC ring of $13\ \mu$s, this corresponds to 2.7 hours and 9 minutes cooling time, respectively. For the two-stage case, $l_{\rm opt}\approx1.0$ for $r=0.2$ so the length of the amplification section is $L=l\sqrt{I_A/I_e}\Sigma\gamma^{3/2}\approx83$ m.

Using the results of the previous section, we can also estimate the diffusion and saturation effects for the eRHIC parameters. For the diffusion caused by the noise in the proton beam, from Eqs.~\eqref{eq:57}, we find the ratio $r_1\approx0.90$, and for the diffusion due to the noise in the electron beam,  Eq.~\eqref{eq:68} yields $r_2\approx8\times10^{-2}$. While both conditions, $r_1,r_2 < 1$ are satisfied, the margin for $r_1$ is not large. From Eq.~\eqref{eq:72} we also find $I_{\rm sat}^{\rm max}\sim 0.85$ which means that in this regime the nonlinear effects are essential. Thus, for the 9 min cooling time, the hadron diffusion and saturation neglected in our study are considerable. The situation can be mitigated by choosing a smaller chicane strength and slower cooling rate. For instance, using $q=0.3$, we obtain a cooling time of 50 minutes, with $r_1\approx0.11$, $r_2 \approx 4 \times 10^{-2}$ and $I_{\rm sat}^{\rm max}\sim0.15$. For these set of parameters, the linear theory of this paper provides a good approximation to reality.

A relatively small value $q=0.3$ also helps with the cooling of hadrons at the tail of the beam energy distribution. As was discussed in Section~\ref{sec:6-1}, for the particles that are shifted longitudinally more than $3.1\Sigma/\gamma$, the effective wake function changes sign and their energy spread increases with time (the anti-cooling effect). With $q=0.3$, the strength of the hadron chicane is $R_{56}^{(h)} = 0.3 \Sigma/\gamma\sigma_h$, and such particles lie at the far tail of the energy distribution, $\eta>10\sigma_h$, where their effect can be neglected.

%
\section{Summary}\label{sec:10}
%
In this paper, we derived the cooling rate for the longitudinal, or momentum, cooling using a simple 1D model that treats particles as charged disks interacting through the Coulomb force. Extending analysis of Ref.~\cite{Stupakov:2018anp}, we studied the cooling with one and two amplification sections in the system. In contrast to Ref.~\cite{Stupakov:2018anp}, where the noise effects are small, adding one or two cascades of signal amplification through a quarter of plasma wavelength drifts and chicanes also amplifies the noise. We have analyzed the role of the diffusion caused by the amplified noise in the electron beam. We also derived formulas that allow estimation of nonlinear effects in the amplification. These effects limit the maximum amplification level that can be used in an MBEC cooling device.

In our analysis, we assumed a round cross section of the beams with a Gaussian radial density distribution. This assumption can be easily dropped and other transverse distributions (e.g., with unequal vertical and horizontal sizes) used for the particle interaction. This will only change the specific form of the interaction potential~\eqref{eq:47}, with the rest of the calculations of the cooling rate remaining the same. 

There are several effects that are neglected in our model. Clearly, the transverse dynamics due to the beam focusing is ignored, as well as longitudinal displacement of particles due to this focusing. We also ignored plasma oscillations in the electron beam in the modulator and the kicker regions. This is justified if the length of the modulator and the kicker is smaller than a quarter of the plasma period in the electron beam. 

Finally, we note that the 1D theory can also be extended to include the effects of the transverse cooling. This type of cooling is achieved through the introduction of the dispersion in the modulator and the kicker regions, as it was proposed for the optical stochastic cooling scheme~\cite{OSC_RHIC,Lebedev:2014cha}. A preliminary consideration of the horizontal emittance cooling in MBEC has been carried out in Ref.~\cite{baxevanis18_s}.

%
\section{Acknowledgments}\label{sec:11}
%

We would like to thank F. Willeke for stimulating discussions of the subject of this work. This work was supported by the Department of Energy, Contract No. DE-AC02-76SF00515.

%

\end{document}